# The G332 molecular cloud ring: I. Morphology and physical characteristics

Domenico Romano,[1,2]★ Michael G. Burton,[1,3] Michael C. B. Ashley,[1] Sergio Molinari,[2] David Rebolledo,[4,5] Catherine Braiding[1] and Eugenio Schisano[2,6]

[1]*School of Physics, The University of New South Wales, Sydney, NSW 2052, Australia*
[2]*Instituto Nazionale di Astrofisica, INAF, Roma Tor Vergata, Via Fosso del Cavaliere 100, I-00133 Roma, Italy*
[3]*Armagh Observatory and Planetarium, College Hill, Armagh BT61 9DG, UK*
[4]*Joint ALMA Observatory, Alonso de Córdova 3107, Vitacura, Santiago, Chile*
[5]*National Radio Astronomy Observatory, 520 Edgemont Road, Charlottesville, VA 22903, USA*
[6]*Inaf - Istituto di RadioAstronomia and Italian ALMA Regional Centre, Via Gobetto 101, I-40129 Bologna, Italy*

**ABSTRACT**

We present a morphological and physical analysis of a giant molecular cloud (GMC) using the carbon monoxide isotopologues ($^{12}$CO, $^{13}$CO, C$^{18}$O $^3P_2 \to {}^3P_1$) survey of the Galactic Plane (*Mopra CO Southern Galactic Plane Survey*), supplemented with neutral carbon maps from the HEAT telescope in Antarctica. The GMC structure (hereinafter the `ring`) covers the sky region $332° < \ell < 333°$ and $b = \pm 0.5°$ (hereinafter the `G332 region`). The mass of the ring and its distance are determined to be $\sim 2 \times 10^5$ M$_\odot$ and $\sim 3.7$ kpc from the Sun, respectively. The *dark molecular gas* fraction – estimated from the $^{13}$CO and [C I] lines – is $\sim 17$ per cent for a CO $T_{ex}$ between [10,20 K]. Comparing the [C I] integrated intensity and N(H$_2$) traced by $^{13}$CO and $^{12}$CO, we define an $X_{CI}^{809}$ factor, analogous to the usual $X_{co}$, through the [C I] line. $X_{CI}^{809}$ ranges between $[1.8, 2.0] \times 10^{21}$ cm$^{-2}$ K$^{-1}$ km$^{-1}$ s. We examined local variation in $X_{co}$ and $T_{ex}$ across the cloud, and find in regions where the star formation activity is not in an advanced state, an increase in the mean and dispersion of the $X_{co}$ factor as the excitation temperature decreases. We present a catalogue of C$^{18}$O clumps within the cloud, and report their physical characteristics. The star formation (SF) activity ongoing in the cloud shows a correlation with $T_{ex}$, [C I], and CO emissions, and anticorrelation with $X_{co}$, suggesting a North–South spatial gradient in the SF activity. We describe virtual reality and augmented reality data visualization techniques, which open new perspectives in the analysis of radio astronomy data.

## 1 INTRODUCTION

In our Galaxy, as in other galaxies, the space between the stars is filled by a faint medium that consists of ionized, atomic, and molecular gas, high-energy particles (cosmic rays), and dust grains at micron scales. Cosmic rays (CRs), magnetic fields, and the turbulent motion of diffuse gas and dust grains, perturb the interstellar medium (ISM) and contribute to it being inhomogeneously distributed at small scales. The molecular structures within the ISM span a wide range of sizes and masses, from Bok globules (Bok & Reilly 1947) of few solar masses on parsec scales (Clemens, Yun & Heyer 1991), where low-mass stars are born (Yun & Clemens 1990), to immense giant molecular clouds (GMCs) with masses of the order of $10^4$–$10^6$ M$_\odot$ spanning tens of parsecs (Bally 1986).

Molecular clouds are the densest and coldest regions of the ISM. Inside them, in regions where CRs, magnetic fields, and violent motions of gas are unable to provide support against gravitational collapse, the molecular gas condenses to form new stars. The ionization and turbulent feedback from the massive newborn stars photodissociates and perturbs the surrounding gas, dissolving the hosting molecular cloud as between 2 and 20 per cent of the original cloud mass is converted into stars (Molinari et al. 2014).

The gas processed by the thermonuclear reactions in the interior of the stars is recycled back into the ISM through mechanisms such as stellar winds and supernova explosions. The injection of matter and energy into the surrounding medium closes the *Galactic ecology* cycle between the ISM and the stars, continuously enriching the ISM gas reservoir with heavy elements and more complex molecules ready to form a new generation of stars.

★ E-mail: domenico.romano@student.unsw.edu.au

The typical time-scale of this cycle is about $5 – 10 \times 10^7$ yr. With the mean star formation efficiency in GMCs being about 5 per cent, a molecule could exist in the ISM for $1 – 2 \times 10^9$ yr before being part of a star (Molinari et al. 2014).

The detection of molecular hydrogen (Carruthers 1970) in the far-UV spectrum of a hot star, and the detection of CO (Wilson, Jefferts & Penzias 1970) as the second most abundant molecule in space, started the era of systematic observation of the distribution of interstellar molecules. Although molecular hydrogen is ubiquitously present in the ISM, and is the primary component of GMCs, it is extremely difficult to directly observe. Its lowest transition energy (quadrupole $J = 2 \rightarrow 0$ line at 28.22 μm) is at 510 K above ground (Dabrowski 1984), very far from the typical GMCs temperatures (10–20 K; Goldsmith 1987). The carbon monoxide molecule, having an almost constant relative abundance with respect to $H_2$ (equal to $\sim 10^{-4}$), and a $J = 1 \rightarrow 0$ dipole transition at 5 K above the ground level, is a good proxy for $H_2$ in the environments of interstellar clouds (Scoville & Sanders 1987). The empirical link between the measured quantity, the $^{12}$CO line flux, and the $H_2$ column density is the $X_{co}$ conversion factor (Dame, Hartmann & Thaddeus 2001; Shetty et al. 2011; Bolatto, Wolfire & Leroy 2013) which is typically $X_{co} \approx$ 2.0–2.7×$10^{20}$ cm$^{-2}$ K$^{-1}$ km$^{-1}$ s for $|b| < 1°$.

Many experimental observations of the ISM distribution across the Galactic Plane have been made and include the emission in various bands of CO (Heyer et al. 1998; Dame et al. 2001; Mizuno & Fukui 2004; Jackson et al. 2006; Burton et al. 2013; Jones et al. 2013; Barnes et al. 2015; Schuller et al. 2017; Umemoto et al. 2017), H I 21cm (Gibson et al. 2000; McClure-Griffiths et al. 2001; Stil et al. 2006), methanol masers (Green et al. 2009), HCO$^+$, N$_2$H$^+$, NH$_3$, water maser, and other tracers (Jones et al. 2012; Jackson et al. 2013; Walsh et al. 2014), 1.3 mm continuum (Aguirre et al. 2011), 870 μm continuum (Schuller et al. 2009), 350–$10^4$ μm Planck (Planck Collaboration 2014), 70–500 μm Herschel (Molinari et al. 2010), 65–160 μm Akari (Ishihara et al. 2010), 3.6–24 μm Spitzer (Benjamin et al. 2003; Carey et al. 2009; Gutermuth & Heyer 2015), 3.4–22 μm Wise (Wright et al. 2010), and 8.2–21.3 μm MSX (Price et al. 2001). The inventory of molecules detected in the ISM ranges from simple ones such as H$_2$O, to organic and more complex molecules such as aldehydes, fullerenes, and polycyclic aromatic hydrocarbon (PAH; Tielens 2013). The ISM is far from being a simple medium.

In the analysis performed in this paper, we will examine in detail the general physical characteristics and the morphology of a GMC recently discovered in the G332 region of the Galactic Plane, identified using data from the Mopra CO Southern Galactic Plane Survey (Braiding et al. 2018).

## 1.1 Paper structure

In this section, we will give a brief description of how we structured our study. We start with a short presentation of our CO data and the relative reduction pipeline (Section 2), followed by an overview of the CO emission in the G332 sector and the molecular ring cloud (Section 3.1), including how we inferred the cloud distance by solving the kinematic distance (KDA; Section 4). In Section 5.1, we describe the adopted data masking method for creating moment maps (Section 5.1), and the derivation of the CO column density (Section 6) assuming local thermodynamic equilibrium (LTE) conditions. The dust spectral energy distribution (SED) fit and its relative molecular hydrogen column density N(H$_2$), relative to the molecular ring cloud, are described in Section 6.5. In order to investigate the local variation of the $X_{co}$ factor, in Section 7.1 we compare the $^{12}$CO integrated intensity map and N(H$_2$) column density derived from $^{13}$CO, at different $T_{ex}$. Taking advantage of the availability of [C I] line data, we define an $X_{CI}^{809}$ factor (Section 7.2) by comparing the molecular hydrogen column density traced by $^{13}$CO and $^{12}$CO to the [C I] integrated intensity. Moreover, we used the atomic carbon emission, in combination with the $^{13}$CO detection, to quantify the dark H$_2$ gas not traced by carbon monoxide (Section 7) and analyse how the [C I] /$^{13}$CO ratio varies in the ring cloud (Section 8.3). To further inspect the different conditions occurring in the cloud, in Section 9 we describe how we identified clumps and report the C$^{18}$O principal physical properties in a catalogue (Table A1). In the discussion section we analyse (Section 10.1) how we disentangled the dust emission across the Galaxy with a spiral arm model, while in Section 10.2 we examine the star formation activity ongoing in the cloud. The relation between the measured $X_{co}$ variation and the physical conditions observed in the cloud are discussed in Section 10.3, and in Sections 10.4 and 10.5, we present ways in which the different physical conditions of the cloud can be inspected using other tracers from CO, such as [C I] and 8μm emission. We end our study (Section 11), presenting new methods based upon virtual reality (VR) and augmented reality (AR) techniques to visualize radio astronomy data, we present three different VR/AR representations in Appendix A.

To help localize the regions of the cloud, single objects, or specific features, we will use the convention shown in Fig. 1. All the global cloud quantities reported in the paper are derived from the region resulting by the union of [SE-R], [SW-R], [W-R], [NW-R], [N-R], and [C-R], if not otherwise stated.

## 2 OBSERVATIONS AND DATA REDUCTION

### 2.1 CO, [C I], and H I

In this paper five spectral lines will be considered: the three CO isotopologues $^{12}$CO, $^{13}$CO, C$^{18}$O $J = 1 \rightarrow 0$ transition (hereinafter called simply $^{12}$CO, $^{13}$CO, and C$^{18}$O), the atomic carbon [C I] $^3P_2 \rightarrow ^3P_1$ transition line at 809.342 GHz (hereinafter [C I]), and the H I 21 cm line.

The CO lines come from the isotopologue survey of $^{12}$CO, $^{13}$CO, and C$^{18}$O, $J = 1 \rightarrow 0$ transition lines (30 arcsec pixel size and $\sim$0.1 km s$^{-1}$ spectral resolution), obtained using the 22 m single dish Mopra millimeter wave telescope, part of the *Mopra Southern Galactic Plane CO Survey* of the Milky Way (Burton et al. 2013, 2014; Braiding et al. 2015; Rebolledo et al. 2016; Blackwell, Burton & Rowell 2017; Braiding et al. 2018). In this survey we used the *fast-on-the-fly* mapping that maps the four CO isotopologues at the same time, taking 8 × 256 ms samples each cycle (2.048 s). In Galactic coordinates it covers $|b| < 1°$ and $265° < \ell < 10°$, each square degree is divided into 20 rectangles of 60 arcmin × 6 arcmin, 10 aligned along lines of Galactic longitude, and 10 aligned with Galactic latitude. Each stripe required $\sim$1 h to be completed, at a scanning speed of $\sim$30 arcsec s$^{-1}$. The CO data presented in this paper, relative to the G332 region, was taken during the first half of 2012 June. The C$^{17}$O line was included in the survey but did not show any detectable emission and will be not considered further.

The [C I] data (72 arcsec beam FWHM, pixel size of 2 arcmin and 0.5 km s$^{-1}$ spectral resolution) comes from the *High Elevation*

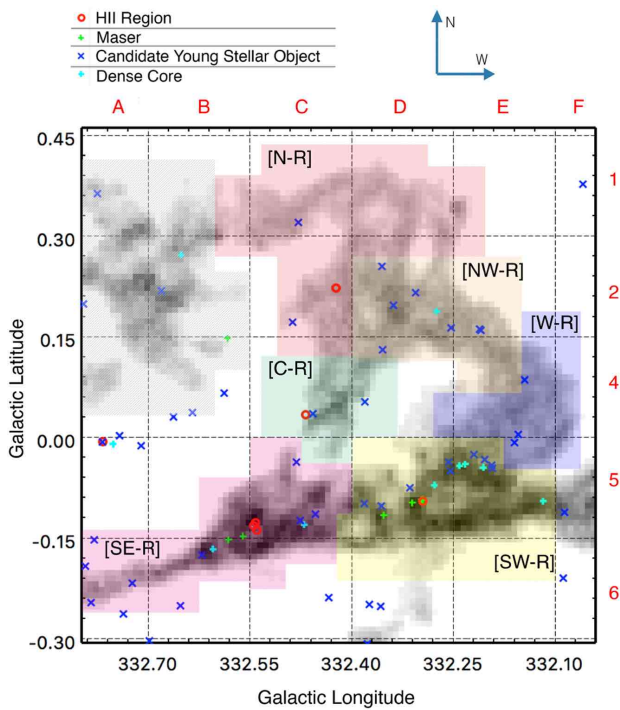

**Figure 1.** Summary of the molecular ring divisions, superimposed on a $^{13}$CO moment zero map, used here to show morphology. We divided the cloud into parts to examine their typical physical conditions: [N-R] light red, [NW-R] in light orange, [W-R] in blue, [SW-R] in yellow, [SE-R] in violet, and [C-R] in green, here the colours are only for identification purpose. In addition, we use the Galactic coordinate grid as reference: from A to F for Galactic longitude and from 1 to 6 for Galactic latitude, both in steps of 0.15°, for a total of 6 × 6 sub-regions. We use the standard astronomical coordinates convention to localize objects inside each sub-sector; for example, the H II region in [N-R] will be indicated as [C2-West]. All the figures in this paper use the Galactic coordinate convention as in the figure above, unless otherwise stated. The region in the upper left, in grey, is a superposition between the North–East extremity of the molecular ring and gas that seems to come from the direction of the RCW106 cloud and is not directly connected to the CO ring. The astrophysical objects are plotted using data from SIMBAD (Wenger et al. 2000) database and DS9 software (http://ds9.si.edu/site/Home.html).

*Antarctic Telescope HEAT*[1] (Kulesa 2011; Burton et al. 2015) located at *Ridge A*, a deep-field site close to the highest dome on the Antarctic plateau, ~1000 km from the South Pole Station.

The H I data come from the Parkes+ATCA Southern Galactic Plane Survey (SGPS, McClure-Griffiths et al. 2005) and has a voxel size of 40 arcsec and a spectral resolution of 0.8 km s$^{-1}$. Table 1 summarizes the characteristics of the surveys we have used.

### 2.2 CO data reduction

Here is given a brief description of the pipeline used to reduce CO data. The raw data in RPFITS[2] format are processed using LIVEDATA[3]

---

[1]HEAT is a 62 cm off-axis Gregorian *FFTS (Fast Fourier Transform Spectrometer)* with heterodyne receivers cooled to 50 K.
[2]http://www.atnf.csiro.au/computing/software/rpfits.html
[3]http://www.atnf.csiro.au/computing/software/livedata/

to subtract the reference position emission. An IDL[4] routine then flags bad data values above a given threshold (usually 5σ). Data are then gridded using the GRIDZILLA[3] package and cleaned again using routines in IDL. The original data reduction pipeline, described in Burton et al. (2013), is used in this work with the only addition of the flagging routine between the LIVEDATA and GRIDZILLA steps. The final output is a square degree FITS datacube with angular dimensions of 131 × 131 pixels, and a velocity range of ($-582$ km s$^{-1}$ to 500 km s$^{-1}$), ($-510$ km s$^{-1}$ to 245 km s$^{-1}$) and ($-515$ km s$^{-1}$ to 215 km s$^{-1}$) for $^{12}$CO, $^{13}$CO, and C$^{18}$O, respectively. All the CO datacubes were then regridded, using the MIRIAD[5] command REGRID[6], to the C$^{18}$O grid, which is (marginally) the coarser spectral resolution compared to $^{12}$CO. The final carbon monoxide datacubes have is composed by voxels an angular voxel dimension of 30 arcsec and a spectral resolution of 0.092 km s$^{-1}$.

### 2.3 Near and far-infrared emission

To correlate our CO data to dust emission we used two surveys of the near-infrared emission coming from the G332 sector: the *Galactic Legacy Infrared Midplane Survey Extraordinaire* GLIMPSE (Benjamin et al. 2003; Churchwell et al. 2009) and the 24μm band MIPSGAL (Carey et al. 2009; Gutermuth & Heyer 2015) observed by the Spitzer space telescope.[7] On the longer wavelength side, to derive the cold dust emission temperature profile, we considered the *Herschel Infrared Galactic Plane Survey* (Hi-GAL, Molinari et al. 2010), which is an Open Time Key Project of the *Herschel* Space Telescope covering the Galactic Plane in a 2° wide strip ($|b| < 1°$) in five photometric bands centred at 70, 160, 250, 350, and 500 μm.

## 3 THE G332 REGION

The H I and $^{13}$CO average emission spectrum coming from the whole G332 square-degree sector is shown in Fig. 2, together with Gaussian curves used to fit the CO emission along the line of sight (see Table 2 for line parameters). The Gaussian curves were obtained using the SCOUSE[8] software (Henshaw et al. 2016).

The right-hand panel in Fig. 2 is a 2D heat map scatter plot[9] between the gas velocity dispersion and the first moment map, computed along the interval $[-115, -10]$ km s$^{-1}$. The pixel distribution shows a branching pattern that is visible at coordinates around $[22, -70]$ km s$^{-1}$ and has nodes at $[12.5, -85]$ km s$^{-1}$ and $[10, -50]$ km s$^{-1}$.

Few pixels have the largest velocity dispersion with velocity centroid in the interval $-60 < V_{\rm LSR} < -80$ km s$^{-1}$, while most of the emission has a second moment below 10 km s$^{-1}$ and a first moment constrained to $-55 < V_{\rm LSR} < -40$ km s$^{-1}$. Small localized $^{12}$CO and $^{13}$CO emissions, missing in the square-degree average profile (Fig. 3), were detected at positive velocities ~5 km s$^{-1}$ and at ~ $-20$, $-7$, and 0 km s$^{-1}$, but this will not be considered in our analysis.

---

[4]http://www.harrisgeospatial.com/SoftwareTechnology/IDL.aspx
[5]http://carma.astro.umd.edu/wiki/index.php/Miriad
[6]https://www.cfa.harvard.edu/sma/miriad/manuals/SMAuguide/smauserhtml/regrid.html
[7]http://irsa.ipac.caltech.edu/Missions/spitzer.html
[8]https://github.com/jdhenshaw/SCOUSE
[9]A 2D heat map is a graph in which scatter plot points are encoded in a discrete grid, each cell of the grid has a different colour representing the number of points inside that cell.

**Table 1.** Parameters for the data sets. The $1\sigma$ rms noise for the CO data was derived from the velocity interval between $-490$ and $-186.5\,\mathrm{km\,s^{-1}}$ in which no emission was detected. For the H I and C I lines we used $-186.3$ to $-135.5\,\mathrm{km\,s^{-1}}$ and $-134$ to $-120\,\mathrm{km\,s^{-1}}$, respectively. For CO the rms is in uncorrected $T_A^*$ K units.

| Survey | Tracer | Frequency (GHz) | Pixel size (arcsec) | Velocity range (km s$^{-1}$) | $\Delta v$ (km s$^{-1}$) | $1\sigma$ (K) rms |
|---|---|---|---|---|---|---|
| MopraCO | $^{12}$CO $J=1$–$0$ | 115.271 | 30 | $-500$ to $+500$ | 0.092 | 1.47 |
|  | $^{13}$CO $J=1$–$0$ | 110.201 | – | $-500$ to $+240$ | – | 0.70 |
|  | C$^{18}$O $J=1$–$0$ | 109.782 | – | $-500$ to $+230$ | – | 0.77 |
| HEAT | C I $J=2$–$1$ | 809.342 | 72 | $-134$ to $+19$ | 0.500 | 0.12 |
| SGPS | H I $S=1$–$0$ | 1.420 | 40 | $-202$ to $+142$ | 0.824 | 1.60 |

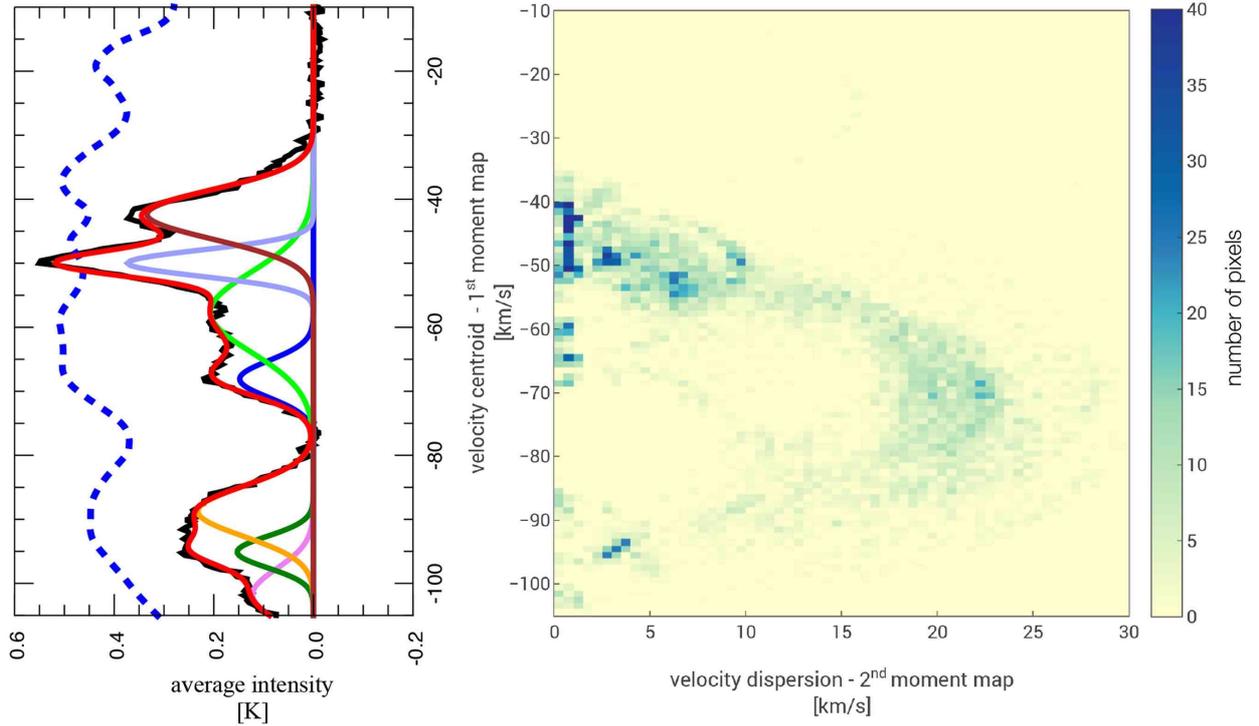

**Figure 2.** *Left-hand image*: Average spectral profile plot along the $V_{\mathrm{LSR}}$ interval $[-105, -10]\,\mathrm{km\,s^{-1}}$ of $T_A^*$ $^{13}$CO (black curve) and H I (blue dashed curve) for the entire G332 square-degree sector, which encloses the region shown in Fig. 1. H I is in $T_{\mathrm{mb}}$ units and reduced by a factor of 200. The red curve is the sum of seven Gaussian curves (displayed in different colours) derived from the multi-Gaussian fit of the $^{13}$CO average emission. See Table 2 for the parameters of each fit. *Right-hand image*: 2D heat map scatter plot comparing the second moment map (velocity dispersion) versus the first moment map (velocity centroid) of each line of sight belonging to the $^{13}$CO datacube, binned by four channels along the velocity axis in the same velocity range as considered in the left-hand panel.

**Table 2.** Parameters for the seven Gaussian fits of the average $^{13}$CO emission in the G332 sector (as shown in Fig. 2). The third column is the *Full Width at Tenth of Maximum* FWTM. The average emission peak is in uncorrected $T_A^*$.

| Velocity centroid (km s$^{-1}$) | FWHM (km s$^{-1}$) | FWTM (km s$^{-1}$) | Peak ($T_A^*$) (K) |
|---|---|---|---|
| $-101.74 \pm 0.54$ | $9.52 \pm 0.82$ | 17.43 | $0.12 \pm 0.01$ |
| $-95.10 \pm 0.20$ | $5.78 \pm 0.58$ | 10.55 | $0.15 \pm 0.03$ |
| $-88.73 \pm 0.32$ | $9.23 \pm 0.47$ | 16.90 | $0.23 \pm 0.01$ |
| $-68.10 \pm 0.15$ | $6.79 \pm 0.57$ | 12.40 | $0.15 \pm 0.02$ |
| $-57.63 \pm 0.31$ | $14.70 \pm 2.53$ | 26.85 | $0.21 \pm 0.01$ |
| $-49.98 \pm 0.04$ | $4.85 \pm 0.16$ | 8.85 | $0.37 \pm 0.02$ |
| $-42.37 \pm 0.14$ | $9.10 \pm 0.19$ | 16.60 | $0.33 \pm 0.01$ |

### 3.1 The G332 giant molecular ring morphology

The average spectral profiles of the CO lines, and [C I], show a peak (left-hand panel of Figs 2 and 3) at $V_{\mathrm{LSR}} = -50\,\mathrm{km\,s^{-1}}$, where a molecular cloud is seen in both atomic and molecular carbon. In order to identify the feature linked to it, we investigated the emission distribution and produced a 3D model[10] of the $^{13}$CO datacube between $-55 < V_{\mathrm{LSR}} < -30\,\mathrm{km\,s^{-1}}$.

---

[10] https://sketchfab.com/models/57aa5cacc17c4fd2b95516d43669b1f8
shows the 3D model of the $^{13}$CO emission represented by two isosurfaces 3D meshes: one at $3.5\sigma$ level and one at $3\,\mathrm{K}$. The mesh grid is 30 arcsec in angular size and $0.35\,\mathrm{km\,s^{-1}}$ in $V_{\mathrm{LSR}}$. In Section 11 we will give a description of how we apply augmented reality to radio data.

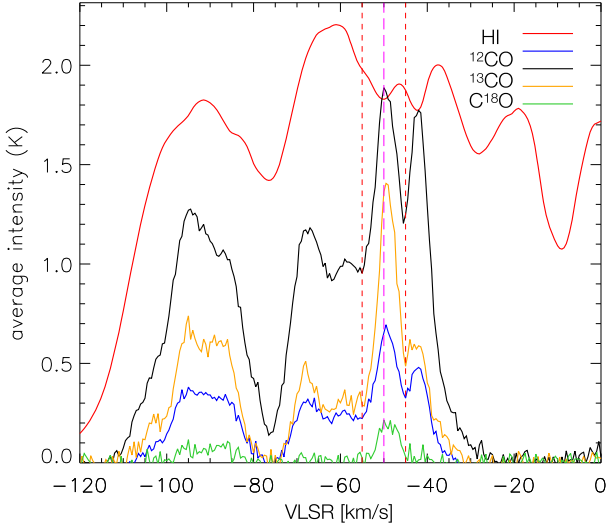

**Figure 3.** Average CO, H I, and [C I] spectral profiles for the ring region shown in Fig. 1. This differs from Fig. 2 in considering the sky region $332.1° < \ell < 332.8°$ and $-0.30° < b < 0.45°$. The two vertical red lines indicate the spectral interval in which the molecular ring is identified. H I and $^{12}$CO are plotted reduced by a factor of 100 and 3, respectively, to allow a clear comparison with other lines. The CO and [C I] peak is at $V_{LSR} \approx -50$ km s$^{-1}$, while, at the same velocity, H I presents a local minimum.

Inspection of the CO spectral line and the sigma threshold masks (see Section 5.1) led us to focus our study on Galactic coordinates $332.0 < \ell < 332.8$, $-0.3 < b < 0.4$, within which we identified the principal emission structure (the *ring*), and on the velocity range $-55 < V_{LSR} < -44$ km s$^{-1}$ (which we refer to as the *ring velocity range*). This choice was made to prevent contamination coming from gas not evidently connected to the ring or with a very low signal-to-noise ratio.[11] The extent defined above is the same used to compute moment maps (as the moment zero in Fig. 4) and all the plots belonging to the ring, unless otherwise specified.

The region delimited by coordinate $\ell > 332.6°$ and $0.4° < b < 0.15°$ (see Fig. 1), where the North–East extremity of the ring and another cloud extremity coming from the East fall in the same ring $V_{LSR}$ range, is excluded from our study, since the two clouds cannot be clearly separated.

Taking into account the above mentioned constraints, the ring angular dimensions are about 0.8° along both South–North and East–West directions, showing mean velocity FWHMs approximately equal to 9, 5, 4, and 8.5 km s$^{-1}$ for the $^{12}$CO, $^{13}$CO, C$^{18}$O, and [C I] average line profiles, respectively.

## 4 THE KINEMATIC DISTANCE AMBIGUITY

We estimated the distance of the ring by using the Galactic rotation model from McClure-Griffiths & Dickey (2007), resolving the KDA by comparing the SGPS HI survey data to the Mopra CO and HEAT surveys. Velocities profiles in Fig. 3 show a local H I minimum in correspondence of CO and [C I] peaks at around $-50$ km s$^{-1}$. Inspecting the CO and H I spectral lines we found a clear decrease in H I where CO peaks, for all of line of sights passing through the ring, indicating the presence of a cold atomic component at the same velocity of the CO peak. This strengthens the interpretation of the ring as a unique connected structure. The near distance solution for $V_{LSR} = -50$ km s$^{-1}$, using a ring geometrical centre $(\ell, b) = (332.3°, 0.0°)$, is $\sim$3.73 kpc from the Sun. Assuming the Galaxy model of Vallée (2014), this distance places the molecular ring inside the Scutum–Crux arm (Fig. 5). For comparison we also used the Bayesian approach[12] of Reid et al. (2016) that gave a distance of $3.22 \pm 0.44$ kpc. However, in our analysis we will adopt the 3.73 kpc value.

---
[11] We note that $^{12}$CO emission in the range $-44 < V_{LSR} < -30$ km s$^{-1}$, constrained in the same ring Galactic coordinates, possesses voxels apparently connected to the ring that are not detected by the threshold masks ($3\sigma$).

In Fig. 6 we present the Galactic spiral arms overplotted with the Position Velocity (PV) image of the $^{12}$CO average emission from the region shown in Fig. 1. It is seen that most of the CO is constrained inside the Norma near, Norma far, and Scutum–Crux near arms, while in the Sagittarius near and Scutum–Crux arms, there are only a few small clouds. The position ambiguity between Scutum–Crux near and Norma far is clearly visible as they are superimposed.

The analysis of the $^{12}$CO spectral lines revealed emission peaks between $-20$ km s$^{-1} < V_{LSR} < 0$ km s$^{-1}$, coming from small weak clouds apparently localized in the Scutum–Crux far arm and in the Sagittarius near arm. Nevertheless, no local H I minimums are present at the same $V_{LSR}$ of these $^{12}$CO peaks, suggesting a far distance solution for their KDA of the order of $\sim$14 kpc.

## 5 MOMENT MAPS

### 5.1 Mask for the moment maps

To improve the detection of the faintest CO emission (as in the case of C$^{18}$O), we used the MIRIAD task IMBIN[13] to bin the original datacubes along the velocity axis by a factor of 4, consequently degrading the spectral resolution to $\sim$0.35 km s$^{-1}$. To separate real emission from noise we applied, on the binned cubes, a custom IDL routine based on the method described by T. M. Dame.[14] Selecting the velocity range, and the spatial region, the extracted cube is 3D smoothed by a boxcar average of a certain width.[15] For CO and [C I] we used a boxcar kernel width of [3,3,2] voxels: the average is computed considering three voxels in each angular direction and two along the spectral axis. The rms noise of the smoothed cube, $\sigma_s$, is used to select only voxels above a $\sigma_{det} = \sigma_s \times \phi_{det}$ threshold. The selection is then expanded including all the surrounding voxels above the $\sigma_{ext} = \sigma_s \times \phi_{ext}$ level. This growing mask procedure is similar to the masking method used in the CPROPS algorithm (Rosolowsky & Leroy 2006). Both the selection and the expansion steps are done on the smoothed cube. All voxels that pass these two criteria form the mask applied to the original cube to compute the moment maps. For $^{12}$CO, $^{13}$CO, and C$^{18}$O the chosen masking threshold values are [6,3], written in the form [$\phi_{det}, \phi_{ext}$]. In this paper we will always refer to this kind of 3D mask coming from a four-channel binned datacube (specifying from which line originates), if not otherwise stated. The same values of smoothing and threshold are used for [C I] and HI. In Fig. 4 are shown the resulting moment maps for CO and [CI]. We opted a boxcar instead of a Gaussian smooth after performing a series of tests applying a custom genetic algorithm (Holland 1992) to find the best combination of the two threshold and two smoothing parameters.

---
[12] http://bessel.vlbi-astrometry.org/bayesian
[13] https://www.cfa.harvard.edu/sma/miriad/manuals/SMAuguide/smauserhtml/imbin.html
[14] https://www.cfa.harvard.edu/rtdc/CO/MomentMasking/
[15] The specific IDL command used is SMOOTH.

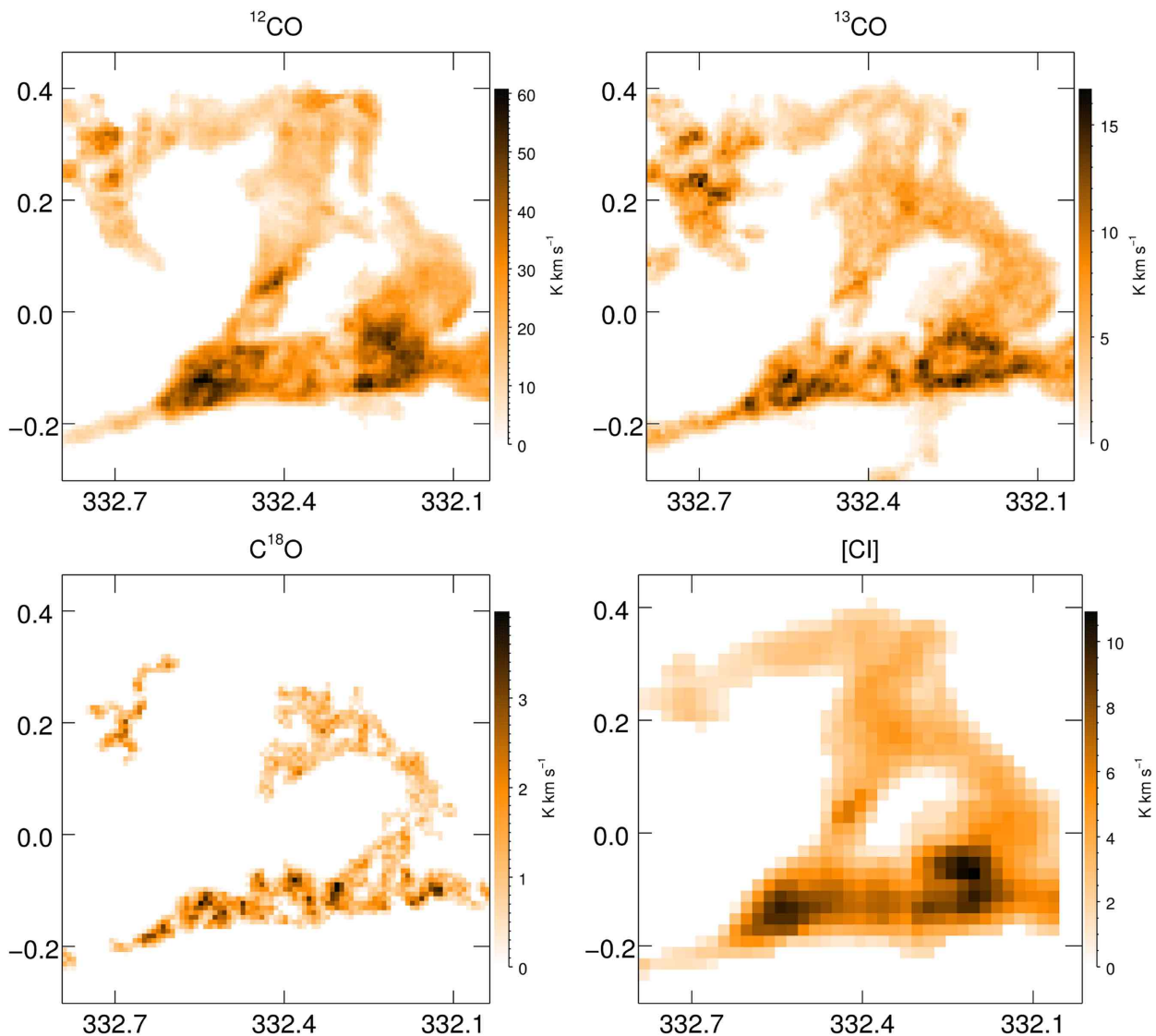

**Figure 4.** Integrated intensity maps of CO lines ($T_A^*$) and C I ($T_{mb}$) over the velocity range $-55\,\mathrm{km\,s^{-1}} < V_{LSR} < -44\,\mathrm{km\,s^{-1}}$. Axes are Galactic with the *x* and *y* indicating longitude and latitude, respectively.

To simplify the description of this algorithm we will take advantage of terms usually used in genetics. The algorithm basically creates a certain number (*population*) of moment map *individuals* from a different choice of the above mentioned four parameters (*four genes or genotype*), randomly selected from a defined range. To each map (*individual*) is assigned a score through a custom defined function (*fitness function*). In our definition the best individuals have the higher fitness function. This function takes into account the number of voxels with negative values included in the 3D mask, the number of voxels of the mask, the number of negative and positive pixels in the final zeroth moment map. In detail, the fitness function is defined as $f_{\mathrm{fit}} = (P^+/1000 - P^-) + (V^+ - V^-)/V_N$, where $P^+$ and $P^-$ are, respectively, the number of positive and negative value pixels in the resulting 2D zeroth moment map, while $V^+$, $V^-$, and $V_N$ are, respectively, the number of positive, negative, and total voxels included in the 3D mask. The major weight given to $P^-$, with respect to $P^+$, is to greatly decrease the fitness function for *individuals* having negative value pixels in the final zeroth moment map. As a consequence, these *individuals* will have minimal chances of passing their *genotype* to the next generation. The parameters (*genotype*) of the elements of the population with the best score are mixed together (*crossover*) to create a new set of parameters (*sons genotype*) introducing a small mutation (*mutation*), thus creating a second generation of the parameter set. When the maximum value of the fitness function does not change for a certain number of iterations, prior to stopping the process, the algorithm increases the variation in the sons parameters (*mutation amplification*) for a maximum number of iterations, in order to prevent the algorithm stopping on a local maximum, exploring a wider parameter space. If a new individual is found the process continues with the mutation rate set to the initial value.

The integrated flux coming from the masked cubes was found to be the 51, 70, and 55 per cent of the integrated flux coming from the non-masked cubes, respectively, for the $^{12}$CO, $^{13}$CO, and C$^{18}$O lines.

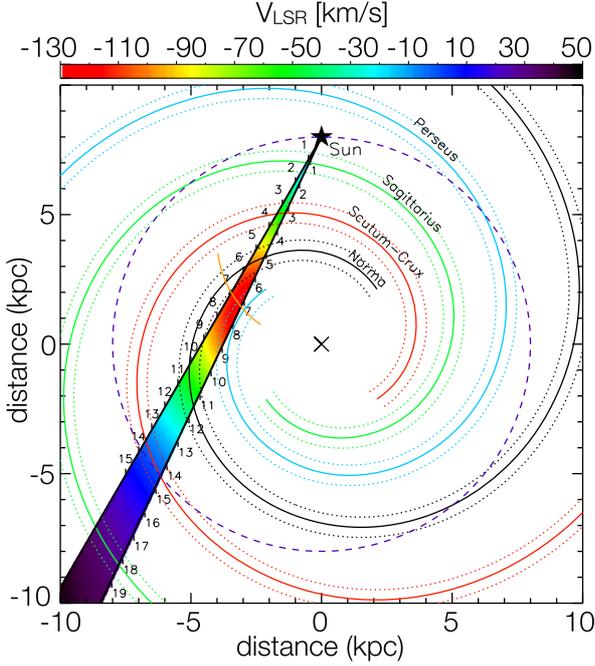

**Figure 5.** Colour representation of the Galactic velocity field for the 330°–335° longitude range, based on the rotation curve and Galaxy model described in Section 4. The numbers along the line of sight indicate the distance from the Sun in kpc. The Solar circle is indicated by the dashed circle while the orange arc indicates the tangent points to the spiral arms.

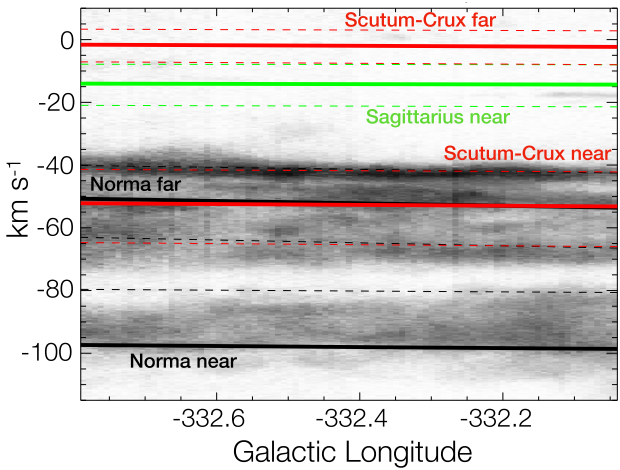

**Figure 6.** PV spiral arms plot, based on the Galactic rotation curve and Galaxy model as described in Section 4, over the $^{12}$CO average intensity PV plot image. The solid lines represent arm mean positions, while segmented lines represent their outer edges. The colour codes for the spiral arms are the same as in Fig. 5.

Comparing the $^{12}$CO and the $^{13}$CO integrated intensity maps, it is possible to see regions in which there is no $^{12}$CO counterpart of $^{13}$CO emission, in particular around $(\ell, b) \sim (332.3°, 0.2°)$. This apparent cavity in [NW-R] is close to PLW2012 G332.277+00.189-048.9 (Purcell et al. 2012), a dense core at $(\ell, b,) \approx (332.274°,$ 0.187°). This is a threshold effect of the 3D mask: in this region $^{12}$CO emission in the smoothed cube is below the masking threshold while $^{13}$CO is above. Examining the line profile (Fig. 7) this behaviour can be explained as $^{12}$CO self-absorption, since the local minimum of the $^{12}$CO spectral lines is close to the $^{13}$CO and C$^{18}$O local peaks. However, a deeper inspection of all the ring $^{12}$CO spectra reveals that this effect is limited to few lines of sight, mainly distributed in the Northern part of the ring, besides the above mentioned region.

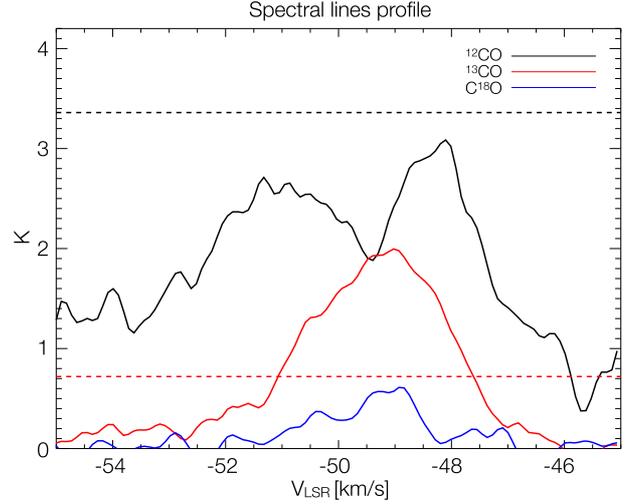

**Figure 7.** Single pixel spectral line profile of the line of sight $(\ell, b) \approx (332.3°, 0.2°)$. $^{12}$CO shows self-absorption, while the $^{13}$CO and C$^{18}$O show a peak close to a $^{12}$CO minimum. In black is $^{12}$CO while in red and blue are $^{13}$CO and C$^{18}$O, respectively. The two dotted horizontal lines represent the $3\sigma$ threshold (black for $^{12}$CO and red for $^{13}$CO) used to derive the mask for the moment maps. The $^{12}$CO is thresholded out even though its emission is stronger than $^{13}$CO.

### 5.2 Integrated intensity

Integrated intensity maps of [C I] and all CO lines are shown in Fig. 4. All the lines show the ring shape, with $^{12}$CO, $^{13}$CO, and [C I] being most similar in shape. Thicker structures seen in C$^{18}$O are distributed along an arc-like structure interrupted by a few cavities. Here we will briefly describe the main features:

(i) $^{12}$CO, $^{13}$CO, and [C I] are higher in [C-R], [SE-R], and [SW-R]. The first region covers a system of multiple bubbles [CPA2006] S52 that encloses [CPA2006] S53, and [SPK2012] MWP1G332437+000547 (Churchwell et al. 2006; Simpson et al. 2012), presenting a bright 8μm emission enclosing diffuse 24μm emission and no significant C$^{18}$O detection.

(ii) [S-R] and [SW-R] host H II regions (Urquhart et al. 2014; see Fig. 1) with velocities compatible with the ring velocity interval: AGAL332.544-00.124 at $V_{LSR} = -46.7$ km s$^{-1}$, close to the highest $^{12}$CO integrated emission and AGAL332.296-00.094 at $V_{LSR} = -48.2$ km s$^{-1}$, near to a $^{13}$CO and a C$^{18}$O integrated emission peak.

(iii) $^{13}$CO and C$^{18}$O maximum integrated emission is at $(\ell, b) \sim (332.40°, -0.09°)$, close to a radio sub-mm source AGAL G332.389-00.091 (Contreras et al. 2013) and a Young Stellar Object (YSO) candidate SSTGLMC G332.3810-00.0992 (Robitaille et al. 2008).

(iv) C$^{18}$O distribution presents a number of dense spots in the southern part, in which there are prestellar and protostellar cores

([SE-R]) as well as dense cores ([SW-R]). The [N-R] shows lower emission, but as in the southern regions there are some YSO candidates (see Fig 1).

The major contribution to the overall ring emission comes from the southern ring that seems to hosts more sites of star formation activity (see Section 10.2) compared to [N-R], [NW-R], and [W-R] regions. As expected, among all the CO lines, $^{12}$CO emission is the most widespread. The apparent absence of $^{12}$CO emission in [NW-R] where $^{13}$CO is present is explained in Section 5.1.

### 5.3 Ring velocity field

The ring velocity field is shown in Fig. 8. The Southern ring emission comes mainly from $V_{\rm LSR} = -50\,\rm km\,s^{-1}$, and is 3 to $-4\,\rm km\,s^{-1}$ wide (see Fig. 9). The Eastern part of [SE-R] has velocity centroid shifts of about $+5\,\rm km\,s^{-1}$ reaching $V_{\rm LSR} = -44\,\rm km\,s^{-1}$ for all the lines, equal to a velocity gradient of $0.30\,\rm km\,s^{-1}\,pc^{-1}$. In the same way, the [C1] region shear is about $0.28\,\rm km\,s^{-1}\,pc^{-1}$.

[N-R] and [W-R] are sheared by about $\sim 2\,\rm km\,s^{-1}$ with respect to the Southern part. Differently from [W-R], in which the velocity field is almost constant at $V_{\rm LSR} \approx -50\,\rm km\,s^{-1}$, [SW-R] has a more complex velocity field, which can be related to the objects in the regions (see Fig. 1).

In [C-R] the first moment map shows a rapid shift of about $5\,\rm km\,s^{-1}$; as said in Section 5.2, this region is associated with many bubbles. Here, $^{12}$CO and $^{13}$CO have broader emissions (see Fig. 9) and show spectral lines in which it is possible to identify two peaks separated by a $\Delta V_{\rm LSR} \approx 5.5\,\rm km\,s^{-1}$.

At coordinates $(\ell, b) = (332.4°, 0.16°)$ a velocity gradient $\Delta V_{\rm LSR} = 1.5\,\rm km\,s^{-1}$, 3 arcmin in diameter, is only visible in $^{12}$CO and $^{13}$CO emissions, and the [C I] resolution is insufficient to distinguish it. In the literature (Peretto & Fuller 2009) are reported two dark clouds close to this position SDC G332.383+0.182 and G332.419+0.174 and Beichman et al. (1988) an IR source IRAS 16111-5032. In the same position, $^{12}$CO and $^{13}$CO spectral lines indicate a dip in the $^{12}$CO emission in correspondence of a $^{13}$CO peak, a scenario compatible with $^{12}$CO self-absorption.

Comparing the first (Fig. 8) and the second moment maps (Fig. 9) it is possible to identify, for all the spectral lines except C$^{18}$O, sharp value changes in [D4] and [D6]. This can be explained as a partial superimposition of cloud fragments detected at a $V_{\rm LSR}$ greater than the main ring $V_{\rm LSR}$. The second moment maps (Fig. 9) show a pattern in [SW-R] that confirms this interpretation: an increase in the velocity dispersion where the first moment maps are suggesting the superposition between the fragments and the ring emission. The same explanation applies to $^{12}$CO in [D1], where both the first and second moment maps show a sharp change in value. In this case, the $^{12}$CO emission of the ring cloud is extended towards lower velocities.

To better understand the physical connection between these extensions and the main ring, we made use of VR and AR representations[10] (described in Appendix A), which allows us to consider them as part of the ring complex.

## 6 CO AND [C I] COLUMN DENSITY AND OPTICAL DEPTH

In order to take in account the response of the Mopra radio telescope, the measured signal ($T_A^*$) was converted into the corrected brightness temperature $T_{\rm MB}$, using the extended beam efficiency $\eta_{xb} = 0.55$ at 115 GHz (Ladd et al. 2005), since the sources considered in this work are more extended than the Mopra main beam. The HEAT datacube is in $T_{\rm mb}$. In Fig. A1 we report the column density maps derived for $^{13}$CO, C$^{18}$O, and [C I] using different $T_{\rm ex}$ values. Our approach to derive $^{12}$CO, $^{13}$CO, C$^{18}$O, and [C I] column densities is based on assuming LTE conditions, and same LTE temperature for all the CO isotopologues. It is important to note that the latter assumption is not verified in the presence of temperature gradients.

Besides LTE conditions, we considered a beam filling factor equal to unity, optically thick $^{12}$CO, and optically thin $^{13}$CO and C$^{18}$O emissions. The last hypothesis was verified by inspecting voxel-by-voxel the C$^{18}$O/$^{13}$CO ratio, which was found to be in the range [0.2–0.25] in regions where the C$^{18}$O is bright enough to yield a significant ratio value within the noise level. It is important to note that this ratio depends also on the oxygen $^{18}$O abundance relative to $^{16}$O as well.

In the next paragraphs we describe in detail all the steps followed to compute optical depth and column densities.

### 6.1 $^{13}$CO and C$^{18}$O

Following the literature, it is possible to define the main brightness temperature ($T_{\rm mb}$) coming from a molecular cloud by the equation of radiative transfer:

$$T_{\rm MB} = f[J(T_{\rm ex}) - J(T_{\rm cmb})][1 - e^{-\tau}], \quad (1)$$

where $\tau$ is optical depth of the considered emission line from the cloud, $T_{\rm ex}$ is the gas excitation temperature, $f$ is the beam filling factor, and $J(T)$ is the Rayleigh–Jeans equivalent temperature, defined as

$$J(T) = \frac{T_0}{e^{T_0/T} - 1}, \quad (2)$$

where $T_0$ is the energy level of the considered transition (5.27, 5.29, and 5.53 K, respectively, for C$^{18}$O, $^{13}$CO, and $^{12}$CO $J = 1 \to 0$ line), and $T_{\rm cmb} = 2.7\,\rm K$. Substituting equation (2) in equation (1), rewriting for $T_{\rm ex}$, and considering optically thick condition for $^{12}$CO ($\tau > > 1$) so that $(1 - e^{-\tau}) \approx 1$, the excitation temperature can be expressed (Wilson, Kristen & Susanne 2009; Mangum & Shirley 2015) by

$$T_{\rm ex} = \frac{T_0^{12}}{\ln\left(1 + \frac{T_0^{12}}{T_{\rm MB}^{12} + J(T_{\rm cmb})}\right)}. \quad (3)$$

Here $T_0^{12} = h\nu_0^{12}/k = 5.53\,\rm K$ and $J(T_{\rm cmb}^{12}) = 0.82\,\rm K$. To better characterize the physical conditions along the ring, we followed two approaches in the derivation of excitation temperature. In the first, we considered a constant $T_{\rm ex}$ for the whole ring, derived from the $^{12}$CO line, using different values of $T_{\rm MB}^{12}$ ranging from 10 to 30 K in steps of 10 K (good approximations to the minimum, median, and maximum main $^{12}$CO brightness temperature in the ring velocity range). The choice of a constant $T_{\rm ex}$ was adopted to verify that a change in the $T_{\rm ex}$ (e.g. a factor of 2, from 10 to 20 K) does not greatly affect the resulting column density. In the second approach, we computed a specific $T_{\rm ex}$ for each line of sight in two different ways: from the $T_{\rm MB}^{12}$ peak map (Fig. 10) and from the dust temperature map (see Section 6.5 and Fig. 11). We note that in the case of $T_{\rm ex} = 10\,\rm K$, for few voxels the optical depth equation (equation 4) has no real solution. These voxels are in correspondence of the warmest region of the cloud, around $(\ell, b) = (332.55°, -0.12°)$ where there are H II regions (see Section 5.2), strengthening the suggestion that here $T_{\rm ex} = 10\,\rm K$ is underestimated.

Rewriting equation (1) for the $^{13}$CO main brightness temperature $T_{\rm MB}^{13}$ in the optically thin regime $(1 - e^{-\tau} \approx \tau)$, substituting $T_{\rm ex}$ and

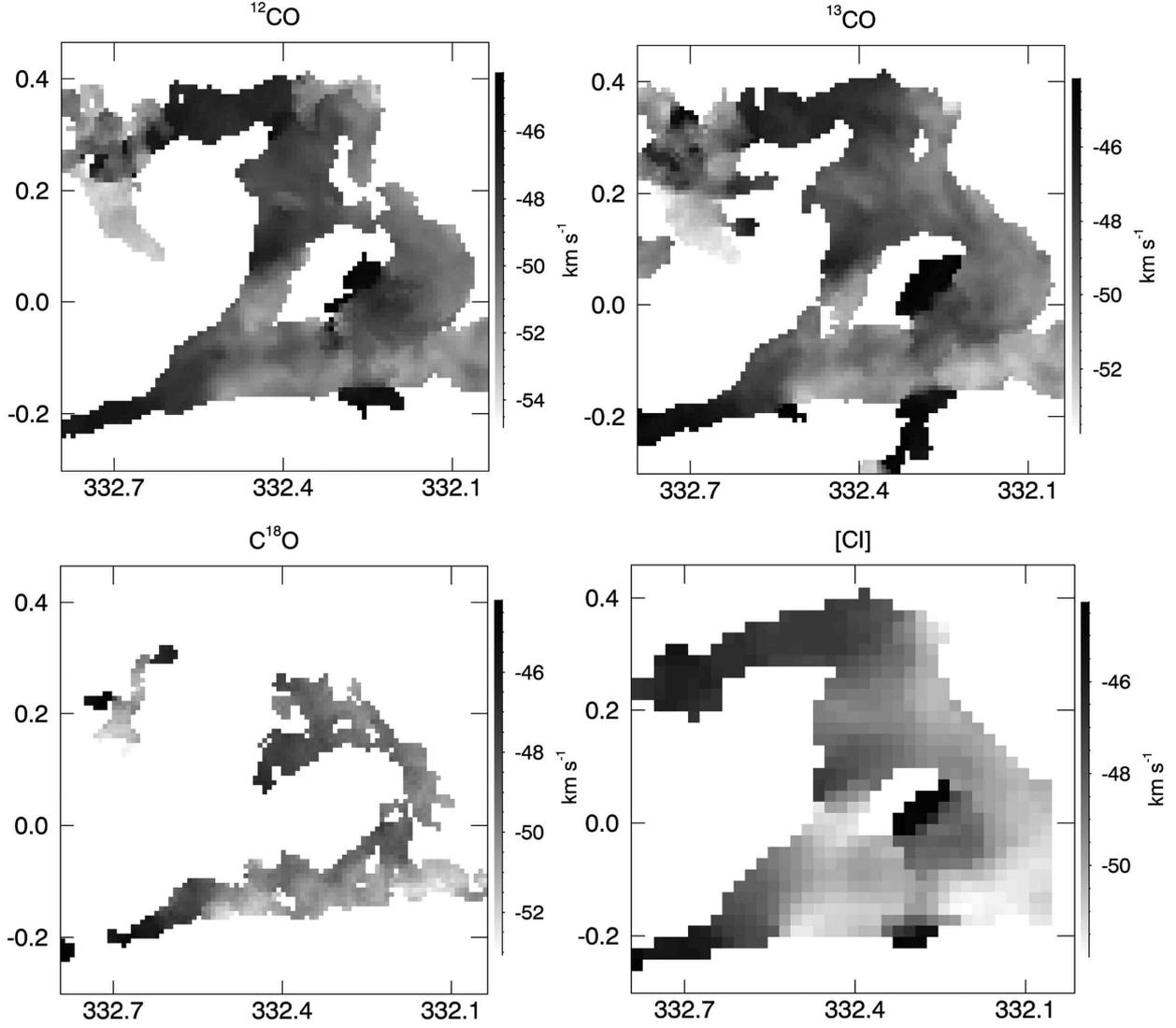

**Figure 8.** Velocity centroid maps of the emission for each line derived from the first moment map.

solving for $\tau$ we obtain

$$\tau_{13} = -\ln\left\{1 - \frac{T_{MB}^{13}}{5.29}\left[\left[e^{5.29/T_{ex}} - 1\right]^{-1} - 0.16\right]^{-1}\right\}. \quad (4)$$

Thus, the $^{13}$CO column density is

$$N(^{13}CO) = 3.0 \times 10^{14} \frac{T_{ex}}{1 - e^{-(T_0^{13}/T_{ex})}} \int \tau_{13} dv, \quad (5)$$

in units of cm$^{-2}$. Temperatures are in K and $dv$ is the velocity interval in km s$^{-1}$. The H$_2$ column density traced by the $^{13}$CO was derived using the ratio [H$_2$/$^{13}$CO] = $5.8 \times 10^5$. This value corresponds to an abundance ratio [H$_2$/$^{12}$CO] = $1.1 \times 10^4$ (Frerking, Langer & Wilson 1982) multiplied by an isotopic abundance ratio [$^{12}$CO/$^{13}$CO] = 53. The latter value was calculated using the (Milam et al. 2005) relation

$$[^{12}CO/^{13}CO] = 6.21 \cdot D_{GC} + 18.71, \quad (6)$$

assuming a ring Galactocentric distance of $D_{GC} \sim 5.5$ kpc.

We applied the same approach used for $^{13}$CO to the C$^{18}$O column density and optical depth computation. To derive the relative column density, we adopted the value from Frerking et al. (1982) of [H$_2$/C$^{18}$O] = $5.88 \times 10^6$, an order of magnitude less abundant than $^{13}$CO.

### 6.2 $^{12}$CO to H$_2$ column density

$^{12}$CO integrated intensity and H$_2$ column density are linked by the X$_{co}$ factor (Sanders, Solomon & Scoville 1984) defined as

$$X_{co} = \frac{N_{H_2}}{W_{co}} \text{ cm}^{-2}\text{K}^{-1} \text{ km}^{-1} \text{ s}, \quad (7)$$

where $W_{co} = \int T_{MB}^{12}(\ell, b, v) dv$ is the integrated intensity in a specified velocity range $dv$ and $T_{MB}^{12}$ is the $^{12}$CO brightness temperature at a specified position. In our analysis we assumed, for the X$_{co}$, the mean Galactic value derived by Dame et al. (2001)

$$X_{co} = 2.7 \times 10^{20} \text{ cm}^{-2}\text{K}^{-1} \text{ km}^{-1} \text{ s}. \quad (8)$$

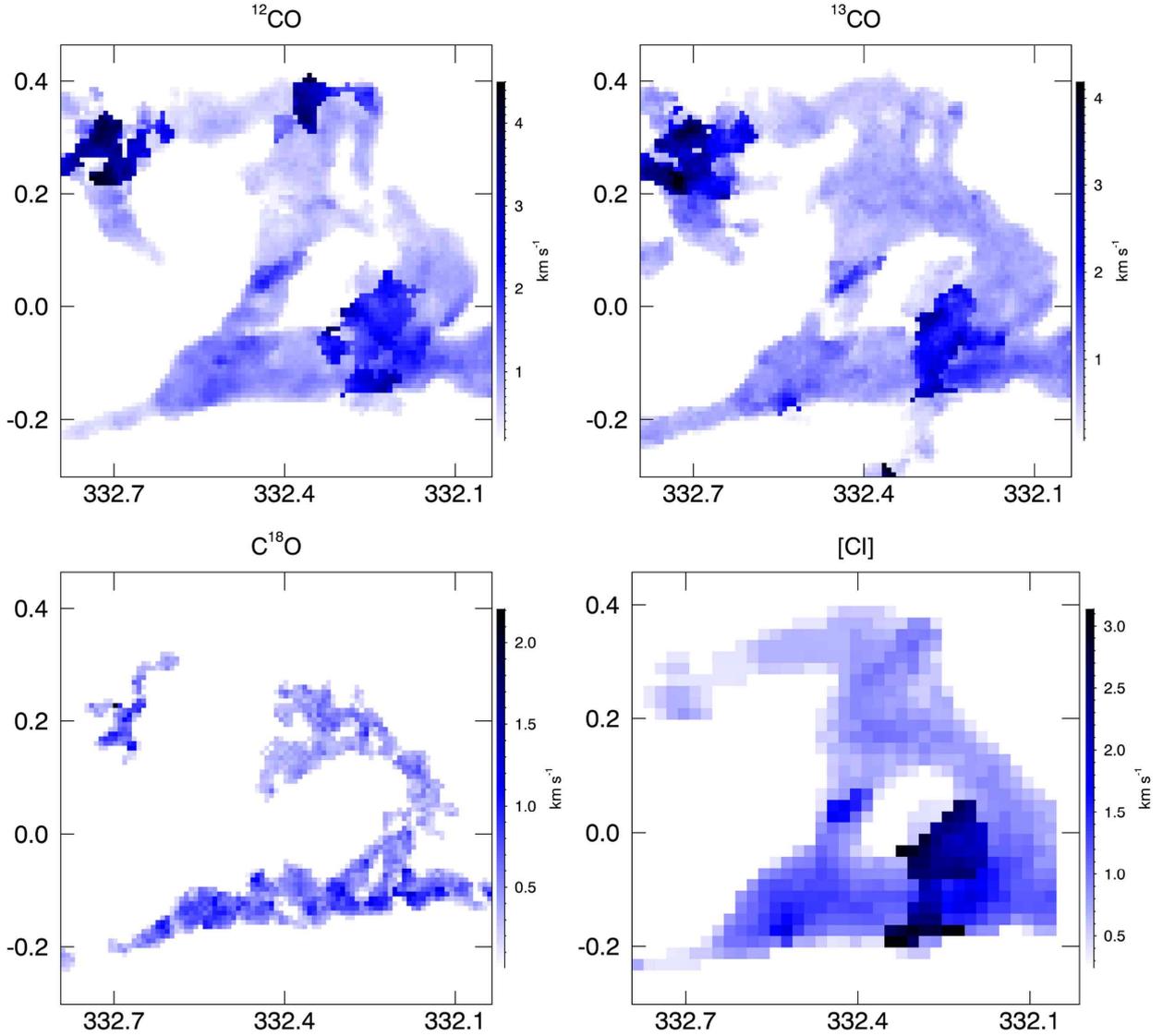

**Figure 9.** Velocity dispersion distribution of the molecular ring for the four emission lines derived from second moment maps.

### 6.3 [C I] $^3P_2 \rightarrow {}^3P_1$

The [C I] column density was derived using the same formulation as in Oka et al. (2001) and Goldsmith & Langer (1999):

$$N_{\rm CI} = \beta_{\rm CI}\left[1 - \frac{J_\nu(T_{\rm cmb})}{J_\nu(T_{\rm ex})}\right]^{-1} Q(T_{\rm ex}) e^{\frac{T_u}{T_{\rm ex}}} \frac{\tau_{\rm CI}}{1 - e^{-\tau_{\rm CI}}}$$
$$\times \int T_{\rm MB}^{\rm CI}(\ell, b, v) dv. \qquad (9)$$

Here

$$\beta_{\rm CI} = \frac{8\pi k \nu_{ul}^2}{g_u A_{ul} hc^3} = 9.5 \times 10^{14}\ {\rm cm}^{-2}\ {\rm km}^{-1}\ {\rm s\ and\ } T_u = 62.46\ {\rm K} \qquad (10)$$

are the values for the [CI] $^3P_2 \rightarrow {}^3P_1$ transition, $T_{\rm MB}^{\rm CI}(\ell, b, v)$ is [C I] voxel brightness temperature, and $A_{ul} = 2.68 \times 10^{-8}$ s$^{-1}$ is the Einstein coefficient for the considered line. The $\tau_{CI}$ is the optical depth correction factor, $\tau_{CI}/(1 - e^{-\tau_{CI}})$, is computed applying equation (4), rewritten for [C I] and using the peak $T_{\rm MB}^{\rm CI}$ value along each line of sight, giving an upper limit on the resulting [C I] column density. The peak $\tau_{CI}$ map for each line of sight calculated assuming a $T_{\rm ex} = 30$ K is shown in Fig. 12. Most of the values are in the range ~0.1 to 0.2 (i.e. optically thin emission). For $T_{\rm ex} = 50$ K, instead, the value of $\tau_{CI}$ decreases by ~60 per cent.

The partition function $Q(T_{\rm ex})$ comes by fitting the values of the *Cologne Database for Molecular Clouds* (CDMS[16]) (Endres et al. 2016) with the following function (Burton et al. 2014):

$$Q(T) = a\left(\frac{T}{1\,K}\right)^b + c, \qquad (11)$$

with *a, b,* and *c* equal to

(i) $a = 16.51$
(ii) $b = 7.12 \times 10^{-2}$
(iii) $c = -17.44$

---

[16] https://www.astro.uni-koeln.de/cdms

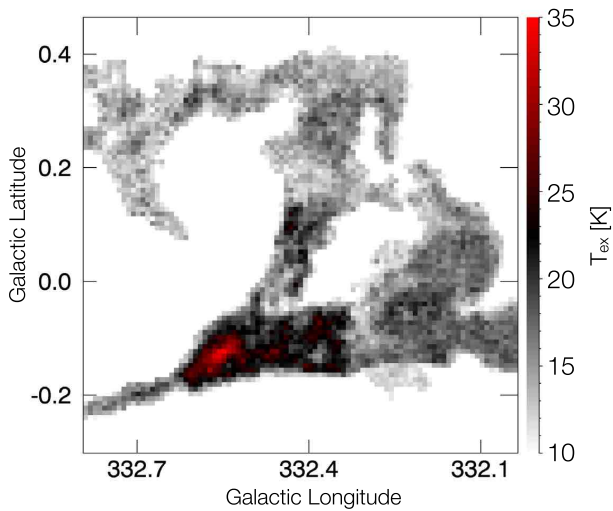

**Figure 10.** Map of $T_{ex}$ as derived from the $^{12}$CO peak emission for each line of sight. The maximum value is ∼35 K.

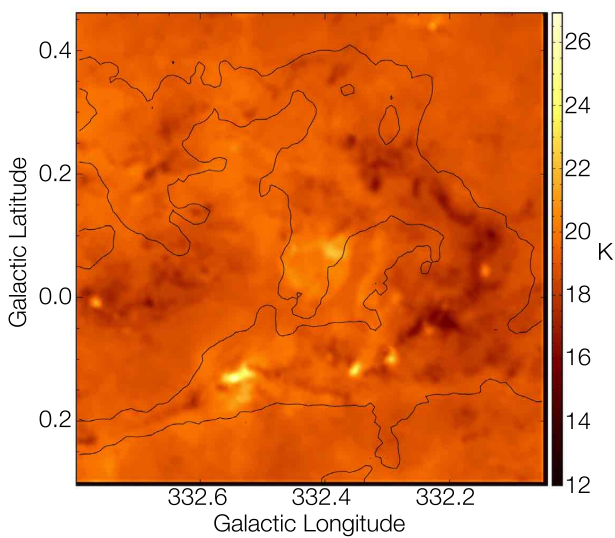

**Figure 11.** Dust temperature map derived from the HiGAL data fit described in Section 10.1. The black contour represents the edge of the $^{13}$CO integrated intensity as plotted in the $^{13}$CO panel of Fig. 4.

We followed the same method used in the CO optical depth calculation shown in equation (4). Differently to CO we choose two a priori excitation temperatures: $T_{ex}$ = 30 and 50 K, both of them higher than the typical CO $T_{mb}$ to take into account that [C I] exists in warmer physical conditions (e.g. the surface regions of molecular clouds, or in the neighbourhood of intense radiation sources). We find comparing to $T_{ex}$ = 50 K, N(CI) increases by a factor of ∼4 and decreases by ∼70 per cent, when using $T_{ex}$ = 20 and 80 K, respectively.

Inspecting the [C I] data we found that the emission is coming from both the external layers around the ring and from the inner part, in a web pattern, in particular in the southern ring. This indicates that localized sources of Far-UV (FUV) radiation are present in the cloud interior. In Section 10.2 we will discuss further the behaviour and interpretation of the [C I] emission.

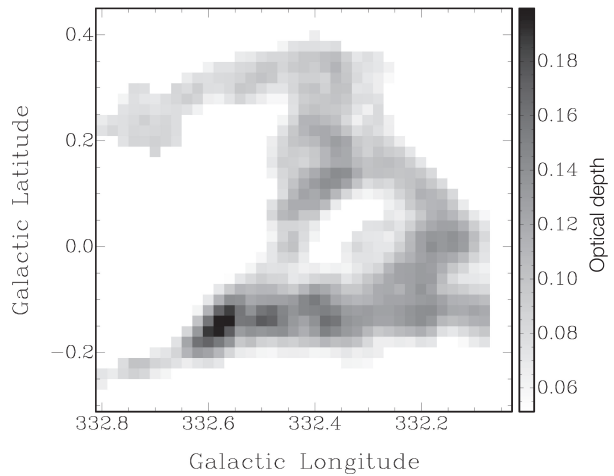

**Figure 12.** Map of [C I] optical depth peak values for each line of sight, derived for a $T_{ex}$ = 30 K. The highest values are in correspondence of the highest $T_{ex}$ derived from the $^{12}$CO peak emission, as shown in Fig. 10, and of the highest dust temperature as plotted in Fig. 11.

**Table 3.** N(H$_2$), CO, and [C I] column density $1\sigma$ sensitivity level for 5 km s$^{-1}$ velocity interval. In the derivation of N(H$_2$) from $^{12}$CO we used $X_{co} = 2.7 \times 10^{20}$ cm$^{-2}$K$^{-1}$ km$^{-1}$ s.

| N(H$_2$) $1\sigma$ sensitivity ($\times 10^{21}$ cm$^{-2}$) | | | |
|---|---|---|---|
| $^{12}$CO | $^{13}$CO | C$^{18}$O | $T_{ex}$ |
| 2.62 | 0.83 | 8.85 | 10 K |
| – | 1.52 | 16.2 | 30 K |

| CO and [C I] column density $1\sigma$ ($\times 10^{15}$ cm$^{-2}$) | | | |
|---|---|---|---|
| $^{13}$CO | C$^{18}$O | C | $T_{ex}$ |
| 1.42 | 1.50 | – | 10 K |
| 2.60 | 2.76 | 6.78 | 30 K |
| – | – | 3.60 | 50 K |

### 6.4 Column density sensitivity

We derived the CO and [C I] column density $1\sigma$ sensitivity by applying the same approach described in Section 6 to an emission-free region 5 km s$^{-1}$ wide (equal to 14 channels in the four-channel binned CO datacube), coherently with the typical spectral width of the lines along the ring. We used a different velocity interval [−130 to −120 km s$^{-1}$] for atomic carbon, since the HEAT datacube has a smaller spectral range than the CO data (see Table 1). The resulting column densities and relative N(H$_2$) $1\sigma$ levels are shown in Table 3 for different $T_{ex}$ values; in Table 4 we report N(H$_2$) quantities derived from different tracers.

### 6.5 Dust spectral energy distribution

In order to retrieve the dust SED we used data from the HiGAL survey (160, 250, 350, and 500 µm). We excluded the 70 µm band

12    *D. Romano et al.*No

**Table 4.** Summary of N(H$_2$) LTE column densities derived from different tracers and from different assumption on the gas temperature, T$_{ex}$. In the first column is indicated the tracer and the corresponding mask: mask 18, mask 13, mask [C I], and mask 12/13 indicates, respectively, the 3D mask created for the moment maps derivation of C$^{18}$O, $^{13}$CO, and [C I] and the intersection between the $^{12}$CO and $^{13}$CO 3D masks (see Section 5.1 for details on mask creation). Mean, max, and sum values are presented for each tracer at different excitation temperatures (sixth column). In particular, T$_{map}$ refers to the T$_{ex}$ derived pixel-by-pixel considering the $^{12}$CO peak emission for each line of sight. The seventh column is the angular size covered by the ring in units of square degrees, while the final column indicates the cloud mass in units of solar masses.

| Tracer | 3D mask | Mean (cm$^{-2}$) | Max (cm$^{-2}$) | Mass surface density (M$_\odot$ pc$^{-2}$) | T$_{ex}$ K | Area (°$^2$) | Mass (M$_\odot$) |
|---|---|---|---|---|---|---|---|
| C$^{18}$O | mask 18 | $1.42 \times 10^{22}$ | $6.51 \times 10^{22}$ | 277 | 10 | 0.08 | $0.94 \times 10^5$ |
| | | $1.87 \times 10^{22}$ | $7.09 \times 10^{22}$ | 375 | 20 | – | $1.27 \times 10^5$ |
| | | $2.48 \times 10^{22}$ | $9.30 \times 10^{22}$ | 484 | 30 | – | $1.64 \times 10^5$ |
| | | $2.01 \times 10^{22}$ | $8.14 \times 10^{22}$ | 392 | T$_{dust}$ | – | $1.33 \times 10^5$ |
| | | $1.85 \times 10^{22}$ | $9.86 \times 10^{22}$ | 345 | T$_{map}$ | – | $1.17 \times 10^5$ |
| $^{13}$CO | mask 13 | $7.93 \times 10^{21}$ | $3.40 \times 10^{22}$ | 166 | 10 | 0.19 | $1.34 \times 10^5$ |
| | | $1.14 \times 10^{22}$ | $4.25 \times 10^{22}$ | 245 | 30 | – | $1.97 \times 10^5$ |
| | | $9.48 \times 10^{21}$ | $3.91 \times 10^{22}$ | 192 | T$_{dust}$ | – | $1.55 \times 10^5$ |
| | | $9.15 \times 10^{21}$ | $4.50 \times 10^{22}$ | 169 | T$_{map}$ | – | $1.36 \times 10^5$ |
| $^{13}$CO | mask 12/13 | $7.88 \times 10^{21}$ | $3.40 \times 10^{22}$ | 153 | 10 | 0.17 | $1.10 \times 10^5$ |
| | | $1.12 \times 10^{22}$ | $4.25 \times 10^{22}$ | 217 | 30 | – | $1.56 \times 10^5$ |
| | | $9.28 \times 10^{21}$ | $3.91 \times 10^{22}$ | 179 | T$_{dust}$ | – | $1.29 \times 10^5$ |
| | | $8.39 \times 10^{21}$ | $4.50 \times 10^{22}$ | 164 | T$_{map}$ | – | $1.18 \times 10^5$ |
| $^{12}$CO | mask 12/13 | $8.39 \times 10^{21}$ | $2.85 \times 10^{22}$ | 162 | – | as mask 12/13 | $1.17 \times 10^5$ |
| Dust | mask 18 | $1.41 \times 10^{22}$ | $5.63 \times 10^{22}$ | 271 | T$_{dust}$ | as mask 18 | $0.92 \times 10^5$ |
| Dust | mask 13 | $1.09 \times 10^{22}$ | | 212 | | as mask 13 | $1.71 \times 10^5$ |
| Dust | mask 12/13 | $1.11 \times 10^{22}$ | | 214 | | as mask 12/13 | $1.54 \times 10^5$ |
| $^{13}$CO | mask [CI] | $6.78 \times 10^{21}$ | $2.75 \times 10^{22}$ | 118 | 10 | 0.25 | $1.25 \times 10^5$ |
| | | $1.03 \times 10^{22}$ | $3.00 \times 10^{22}$ | 180 | 30 | – | $1.91 \times 10^5$ |

since the emission at this wavelength can have significant contributions from non-thermal process (Schneider et al. 2012; Elia et al. 2013). All the maps were convolved to the coarsest resolution of 500 µm (11.5 arcsec pixel size) and placed on the same grid. The pixel-by-pixel fit was done using the modified blackbody function

$$F_\nu = \Omega \left(1 - e^{-\tau_\nu}\right) B_\nu(T_d). \quad (12)$$

Here we assume that the emission at these wavelengths comes from optically thin dust, so that $\left(1 - e^{-\tau_\nu}\right) \approx \tau_\nu$, and a dust optical depth described by the exponential law (Hildebrand 1983)

$$\tau_\nu = \left(\frac{\nu}{\nu_0}\right)^\beta, \quad (13)$$

where $\nu_0$ is the reference frequency and $\beta$ is the dust emissivity index. Since $\tau_\nu = \kappa_\nu n_d$, where $\kappa_\nu$ is the dust opacity and $n_d$ is the mass column density in units of g cm$^{-2}$, we have the same scaling for the dust opacity. The beta parameter depends on the dust grain properties, also showing an inverse correlation with the dust temperature (Dupac et al. 2001). Considering the hypothesis of crystal and metal dust grains we assumed a fixed value of $\beta = 2$ ($\beta = 0$ for a blackbody), following Boulanger et al. (2002). The final form of the fitted function is

$$F_\nu = \Omega \, \kappa_0 \left(\frac{\nu}{\nu_0}\right)^\beta n_d \, B_\nu(T_d). \quad (14)$$

We choose for the dust opacity at 500 µm a value of $\kappa_0 = 3.79$ g$^{-1}$ cm$^2$ (Ossenkopf & Henning 1994) consistent with dust grains with thick ice mantles. The dust temperature fit was performed using the same routine described in Rebolledo et al. (2016). Fig. 11 shows the resulting temperature map. The differences between the molecular gas and dust distributions arise because the dust map reveals all the Galactic emission along the line of sight, crossing multiple spiral arms, whereas the molecular ring emission is constrained within a limited velocity range, coming from a single spiral arm. Nevertheless, it is possible to infer connections between the gas and dust. In Section 10.1 we will consider how to disentangle dust emission coming from the multiple spiral arms using our CO data.

### 6.6 Dust to H$_2$ column density

The derivation of the H$_2$ column density from the dust mass column density $n_d$ (in units of g cm$^2$) was calculated using the relation

$$N^d_{H_2} = \frac{n_d}{\mu_{H_2} m_H R_d}, \quad (15)$$

where $\mu_{H_2} = 2.72$ is the mean molecular weight, $m_H$ is the hydrogen mass in grams, and $R_d = 0.01$ is the dust-to-gas ratio. In this approach we assumed that the gas traced by dust is all in molecular form, and this may not be true in the external layers of the clouds or in high radiation field regions.

The map of N(H$_2$) dust is shown in Fig. 13, overlaid by the $^{13}$CO moment zero map, computed considering the $V_{LSR}$ range [$-115$ to $10$ km s$^{-1}$].

In order to seek a correlation between N(H$_2$) dust and total CO emission along the whole line of sight, in Fig. 14 we shown a scatter plot between these two quantities. Depending on the correlation between N(H$_2$) dust and $^{13}$CO integrated intensity, two groups of points were identified: the red ellipse group (A) possesses an higher correlation compared to the black ellipse group (B). The linear fit overlaid on the scatter plot is computed on group A points only. In doing so, we are assuming N(H$_2$) traced by $^{13}$CO is proportional to

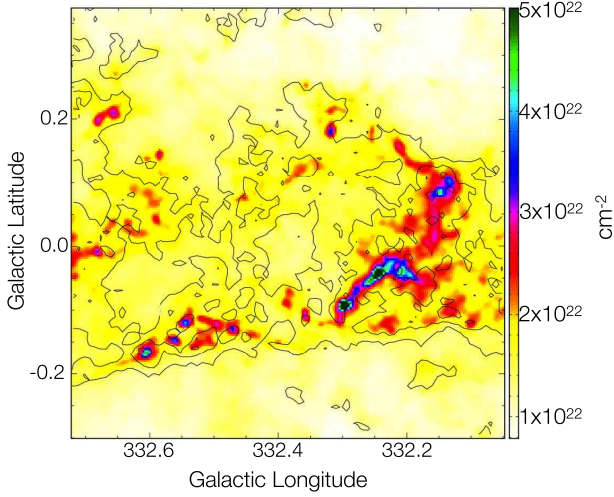

**Figure 13.** H$_2$ column density map derived from the modified blackbody fit to the Hi-GAL data. The black contours represent the $^{13}$CO integrated intensity relative to the V$_{LSR}$ interval [−115 to 10 km s$^{-1}$]. The contours steps are 25, 50, 75, and 100 per cent of the maximum value of 30 K km s$^{-1}$. Scale bar shows the column density in units of cm$^{-2}$.

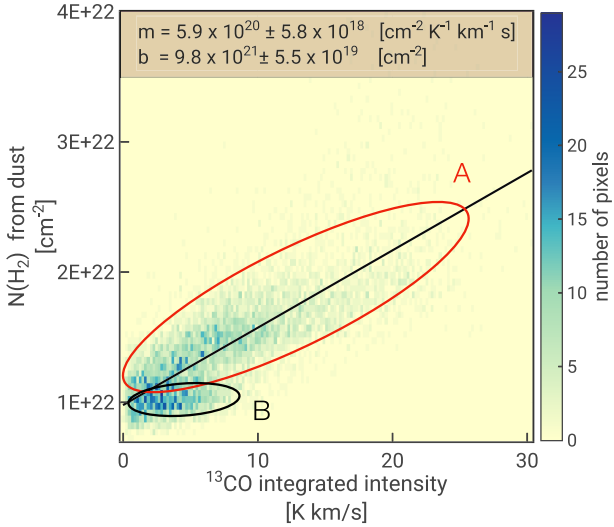

**Figure 14.** 2D heat map scatter plot between N(H$_2$) dust and $^{13}$CO integrated intensity along the whole V$_{LSR}$ emission range [−115 to 10 km s$^{-1}$]. We distinguished two groups of points. Group A (red ellipse) showing a greater correlation between N(H$_2$) dust and $^{13}$CO integrated intensity compared to the group B (black ellipse). They will be further discussed in Section 10.1. The fit line in black is computed only on group A points, fit parameters are shown at the top, N(H$_2$)$_{dust}$ = m · F$_{13CO}$ + b.

$^{13}$CO integrated intensity. We used a 2D heat map representation to show pixel values for density distribution, as shown in the scale bar in Fig. 14 (further discussed in Section 10.1).

### 6.7 Column densities distribution

Logarithmic histograms of N(H$_2$) traced by $^{13}$CO are illustrated in Fig. 15, where the red dashed lines represent the sensitivity limit on the column densities as reported in 3. We used two different values for T$_{ex}$: 10 and 30 K. The fit on these distributions was done considering the histogram values in the interval (21.0, 22.7 Log[cm$^{-2}$]) and (21.2, 22.7 Log[cm$^{-2}$]), using a lognormal function of the form

$$N_{pix} = N_{pix}^{peak} \times \exp\left(-4\text{Log}(2)\frac{(\text{Log}[N(H_2)] - \text{Log}[N(H_2)_p])^2}{\text{FWHM}^2}\right), \quad (16)$$

where N$_{pix}^{peak}$ is the distribution peak, and Log[N(H$_2$)$_p$] is the corresponding N(H$_2$) column density logarithmic value in units of Log[cm$^{-2}$]. Lognormal function parameters, for the distributions in Fig. 15, are shown in Table 5. We note that the fitting functions mostly follow histogram distributions. However, they overestimate column densities for Log[N(H$_2$)] > ∼22.5 and underestimate column densities with values below the sensitivity limit in both cases with T$_{ex}$ = 10 and 30 K.

## 7 [C I] AND *Dark* MOLECULAR GAS

In the surface layer of molecular clouds FUV radiation photodissociates CO to atomic carbon and ionized carbon (C$^+$), whereas H$_2$ is able to self-shield due to its higher abundance; thus the H$_2$ not traced by CO is traced instead by [C I] emission (Wolfire, Hollenbach & McKee 2010). To quantify this dark molecular gas fraction of H$_2$ not traced by CO we compared column densities coming from CO and [C I], N$_{CO}$, and N$_{CI}$, respectively, considering as the dark gas fraction the quantity

$$f_{DG} = N_{CI}/(N_{CO} + N_{CI}). \quad (17)$$

To derive N$_{CO}$ we converted the $^{13}$CO column density into total CO column density applying the same isotopic abundance ratio [$^{12}$CO/$^{13}$CO] = 53 derived in equation (6) for the assumed ring Galactocentric distance of 5.5 kpc. We computed two f$_{DG}$ pixel-by-pixel maps (see Fig. 16) which differ only for the value of T$_{ex}$ (10 or 20 K) used to calculate N$_{CO}$, while for N$_{CI}$ we used a fixed T$_{ex}$ = 30 K. All column densities were derived applying the same 3D mask to both [C I] and CO datacubes. The $^{13}$CO datacube was regridded to the [C I] resolution and then used to compute the 3D mask to compare their emissions.

Examining the images, in Fig. 16 further we see that, in general, the dark gas fraction possesses lower values on lines of sights passing through the innermost part of the molecular cloud, while increasing towards the edges of the CO ring. In particular, the lowest values are located in [NW-R]. This is consistent with an increased fraction of dark gas in molecular cloud surfaces where CO is photodissociated.

Since column density increases almost linearly with the assumed T$_{ex}$ (for excitation temperatures above the energy of the considered emission line), we found lower f$_{DG}$ values for N$_{CO}$ computed at T$_{ex}$ = 20 K than at T$_{ex}$ = 10 K. In particular the average f$_{DG}$ at T$_{ex}$ = 20 K is 16 per cent, while at T$_{ex}$ = 10 K it is 18 per cent. The latter is similar to the f$_{DG}$ computed assuming a T$_{ex}$ derived from the $^{12}$CO peak temperature map (Fig. 10). The molecular dark gas fraction decreases if we instead assume [C I] T$_{ex}$ = 50 K. In this case the average f$_{DG}$ values are equal to 10 per cent for a CO T$_{ex}$ = 10, 20 K, or derived from the $^{12}$CO peak temperature map. Comparing with other estimates, our average values are nearly half of the value estimated in Burton et al. (2015), which gave a value of ∼30 per cent for the average dark molecular gas fraction, for a cold quiescent molecular cloud in G328.

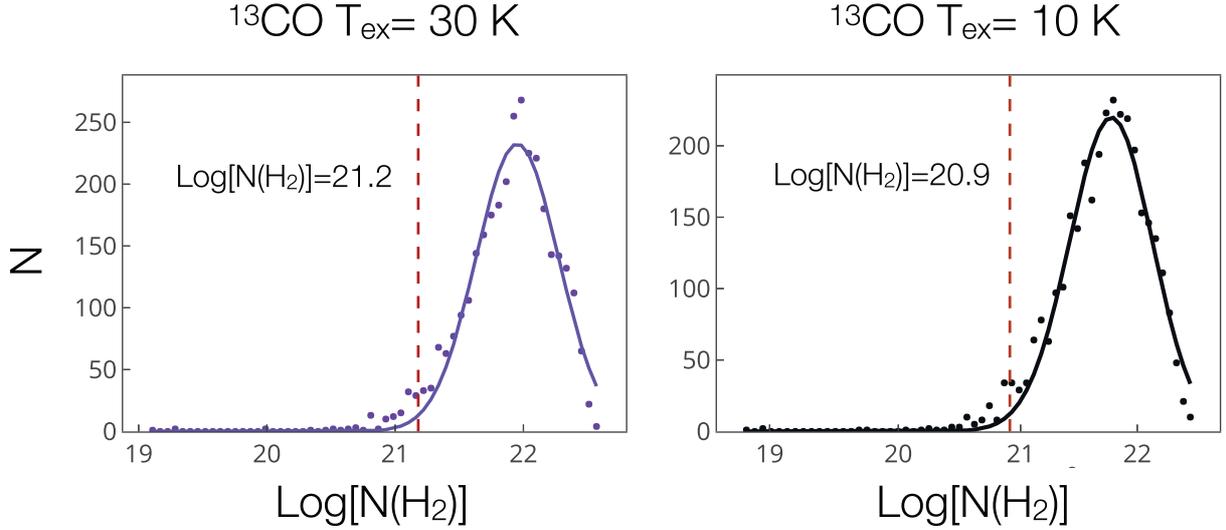

**Figure 15.** N(H$_2$) histograms (*x* axis are in logarithmic scale) of N(H$_2$) column densities traced by $^{13}$CO for different T$_{ex}$ as indicated on each graph. Histograms are derived from voxels belonging to the 3D mask resulting from the intersection between $^{12}$CO and $^{13}$CO relative masks. The red dashed lines indicate the column density sensitivity limits in units of Log[cm$^{-2}$]. Their values are 21.2 and 20.9, respectively, for T$_{ex}$ = 30 and 10 K.

**Table 5.** Summary table of the parameters for the lognormal fits done on the histogram distribution of the column densities derived at different excitation temperatures as shown in Fig. 15.

| | Lognormal fit parameters | | |
|---|---|---|---|
| T$_{ex}$ | Log[N(H$_2$)$_p$] | FWHM | N$_{pix}^{peak}$ |
| K | Log[cm$^{-2}$] | Log[cm$^{-2}$] | – |
| 10 | 21.773 ± 0.012 | 0.845 ± 0.031 | 220 |
| 30 | 21.952 ± 0.013 | 0.756 ± 0.031 | 230 |

### 7.1 X$_{co}$ factor

We inferred the X$_{co}$ factor by comparing the $^{12}$CO integrated intensity map with the N(H$_2$) map obtained from $^{13}$CO computed at different T$_{ex}$. In the derivation of these maps, we adopt for the 3D mask the intersection of the single 3D masks coming from $^{12}$CO and $^{13}$CO (see Section 5.1), in order to ensure comparison of the emission from the same voxels.

In Fig. 17 is shown the ratio between the X$_{co}$ derived from different T$_{ex}$ and the X$_{co}$ computed at T$_{ex}$ = 10 K, indicated as X$_{CO}$(T$_{10}$). In doing so, we used the T$_{ex}$ values of 20, 30, 50 K, and the T$_{ex}$ derived from the $^{12}$CO peak emission map for each line of sight, indicated in the plot as T$_{ex}$ = T$_{map}$. We note that the X$_{co}$ factor is nearly inversely dependent on T$_{ex}$. In particular, taking as reference the X$_{co}$ derived at T$_{ex}$ = 10 K (for which the median value is ≈2.1 × 10$^{20}$ cm$^{-2}$ K$^{-1}$ km$^{-1}$ s), the X$_{co}$ at T$_{ex}$ = 50 K is three times higher at lower X$_{co}$ values while for higher X$_{co}$ the ratio decreases to 2. Further, the scatter increases increasing X$_{co}$. A similar trend is found for all assumed values of T$_{ex}$.

In Fig. 18 we plotted the angular distribution of X$_{co}$ for T$_{ex}$ = T$_{map}$, the median value is 2.2 × 10$^{20}$ cm$^{-2}$ K$^{-1}$ km$^{-1}$ s. Small scale (compared to cloud extension) X$_{co}$ variations are present in almost all the cloud, up to a factor of 2. In general, there is no direct relation between X$_{co}$ and integrated intensity (Fig. 4). An exception is in [C2], [D2], and [E2], where there are evidences of an inverse proportion between X$_{co}$ and $^{12}$CO integrated intensity. This is not unexpected since $^{12}$CO self-absorption in these regions (see last part of Section 5.1).

In Fig. 19 is illustrated the corresponding histogram, showing that the values are mostly (∼80 per cent) in the range between 1.7 × 10$^{20}$ and 3.5 × 10$^{20}$ cm$^{-2}$ K$^{-1}$ km$^{-1}$ s.

A deeper discussion on X$_{co}$ variation across the ring molecular cloud is presented in Section 10.3.

### 7.2 X$_{CI}^{809}$ factor

Similarly as for X$_{co}$, we defined an X$_{CI}^{809}$ factor to link [C I] 809 GHz integrated intensity W$_{CI}^{809}$ to the molecular hydrogen column density

$$X_{CI}^{809} = \frac{N(H_2)}{W_{CI}^{809}}. \quad (18)$$

To obtain X$_{CI}^{809}$ we compared the [C I] integrated intensity with N(H$_2$) derived from $^{12}$CO and from $^{13}$CO at T$_{ex}$ = 10 K. In the latter case we computed two values for N(H$_2$): one adding the dark gas fraction inferred applying equation (17) and one without considering it.

The linear fit m · F$_{CI}$, being F$_{CI}$ the [C I] integrated intensity in K km s$^{-1}$, gives for X$_{CI}^{809}$ the values of (1.77 ± 0.03) × 10$^{21}$, (1.83 ± 0.02) × 10$^{21}$, and (1.97 ± 0.02) × 10$^{21}$ cm$^{-2}$ K$^{-1}$ km$^{-1}$ s, respectively, for N(H$_2$) traced by $^{12}$CO, $^{13}$CO without adding dark gas and $^{13}$CO adding the dark gas fraction.

We report the value of X$_{CI}$ = 1.10 × 10$^{21}$ cm$^{-2}$ K$^{-1}$ km$^{-1}$ s by Offner et al. (2014) and X$_{CI}$ = 1.01 × 10$^{21}$ cm$^{-2}$ K$^{-1}$ km$^{-1}$ s obtained from the numerical simulation of Glover et al. (2015) that were derived for the CI emission at 609 μm, about half of the value we derive from the data.

## 8 CO AND [C I] LINE RATIOS

### 8.1 EVANS plot

In the following, we present ratios between the $^{12}$CO, $^{13}$CO, and [C I] lines using a signal-to-noise (SNR) visual discriminator type of graph known as EVANS[17], where high SNR pixels are more evident than lower ones. Here we give a brief description on how to read EVANS plot taking as reference Fig. 20 (see also Burton et al. 2015). Each plot consists of two pictures, one showing data (left-hand panel) and one showing a Hue Vs Saturation colour table (right-hand panel). Each pixel in the data map has hue and saturation values, which are codified into SNR and data values (the ratio value in this case) using the colour table, according to the axes scales. In this way, those pixels with high SNR values are more evident with respect to lower SNR pixels, which fade into grey independently from their ratio value, giving them a lower visual weight.

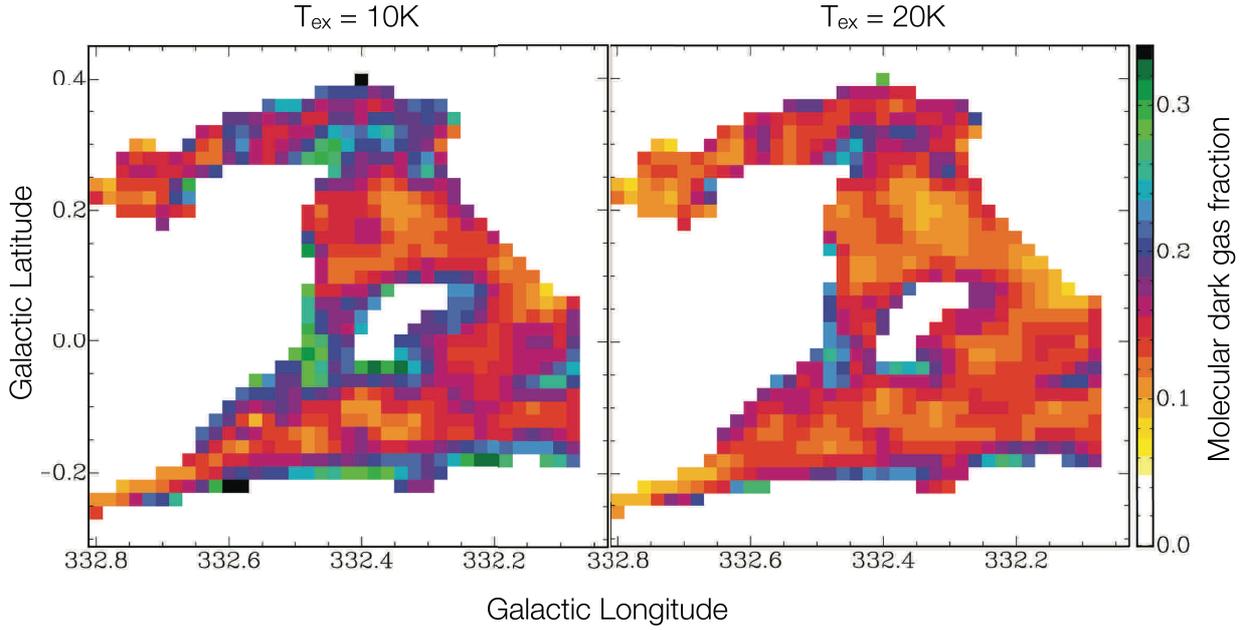

**Figure 16.** *Dark molecular gas* fraction map across the molecular ring. We used a fixed $T_{ex} = 30$ K for the [C I] column density while for CO the used two $T_{ex}$ values as indicated above each panel. The scale is the same for both pictures.

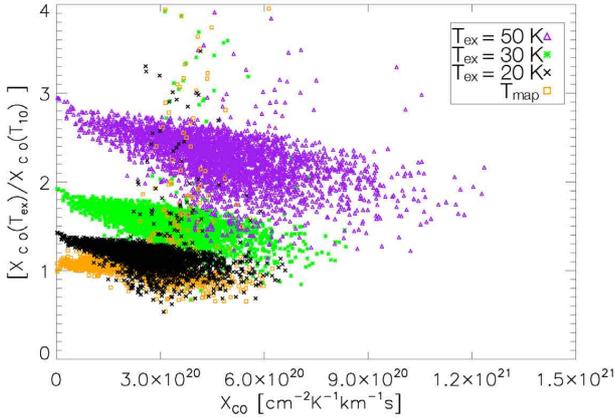

**Figure 17.** $X_{co}$ factor distribution comparison (derived as described in Section 7.1), *y* axis is in the form $X_{co}(T_{ex})/X_{co}(T_{10})$, where the excitation temperature $T_{ex}$ is equal to 20, 30, and 50 K, respectively ($T_{10}$ is for $T_{ex} = 10$K), $T_{10}$ median value is $2.2 \times 10^{20} \text{cm}^{-2} \text{ K}^{-1} \text{ km}^{-1}$ s. $T_{map}$ means that column density calculation was based on the pixel-by-pixel $T_{ex}$ map (Fig. 10), derived for each line of sight from the $^{12}$CO peak emission.

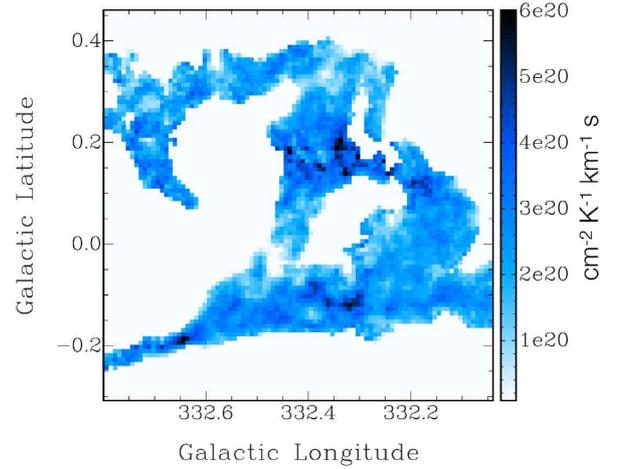

**Figure 18.** $X_{co}$ factor local variation map derived from the ratio between the $^{12}$CO integrated intensity and the N(H$_2$) computed from $^{13}$CO for a $T_{ex}$ derived from the $^{12}$CO peak temperature map.

### 8.2 $^{12}$CO/$^{13}$CO ratio

In Fig. 20 is shown the ratio between integrated intensity map of $^{12}$CO and $^{13}$CO, over the ring spectral channel range. Each zeroth moment map was derived applying the same $^{13}$CO 3D mask to both datacubes. The ratio exhibits almost everywhere a value >1, increasing towards the cloud edge respect to innermost regions, in which it assumes values in the range [2–4]. A small region located in [D2], around $(\ell, b) = (332.3°, 0.2°)$, shows a ratio <1; this could be explained by $^{12}$CO self-absorption (Fig. 7).

[17] https://www.ncl.ucar.edu/Applications/evans.shtml

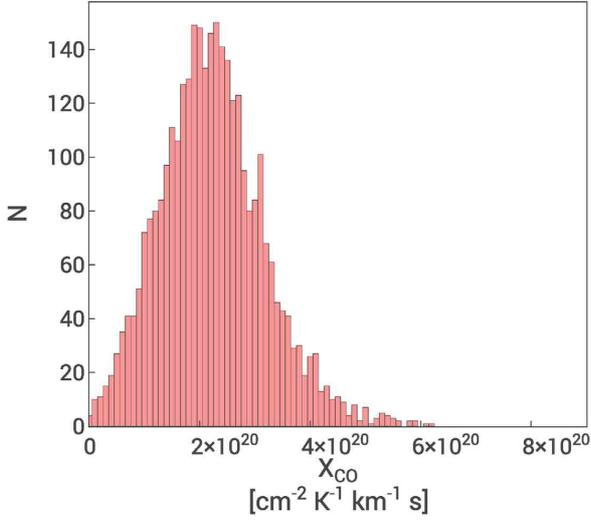

**Figure 19.** Histogram of the $X_{co}$ distribution derived from a $T_{ex}$ computed from $^{12}$CO peak emission map as in Fig. 18. The median value is $\approx 2.2 \times 10^{20}$ cm$^{-2}$ K$^{-1}$ km$^{-1}$ s.

The median value of $R_{12/13} \approx 3.3$, while in the limit of optically thick lines the ratio tends to $\sim 2.6$. Most of the $^{13}$CO emission comes from optically thin gas, with an emission ratio in the range [2–6]. The relation between CO optical depth and brightness temperature ratio is

$$R_{12/13} = \frac{T^{12}}{T^{13}} = \frac{1 - e^{-\tau_{12}}}{1 - e^{-\tau_{13}}}. \quad (19)$$

Assuming the isotope abundance ratio $X_{12/13} = [^{12}C/^{13}C]$ equals to the isotopologue ratio (see equation 6), in the limit of $\tau_{12} > 1$ and $\tau_{13} < 1$ (which is commonly satisfied for most measurements), it is possible to rewrite equation (19) as

$$\tau_{12} \approx \frac{X_{12/13}}{R_{12/13}}. \quad (20)$$

Looking at Fig. 20, the highest $\tau_{12}$ values are in [NW-R], as well as the most uncertain (possessing the lowest saturation values) in correspondence of the lowest ratio values. The southern ring, however, has a less homogeneous range.

### 8.3 [C I]/$^{13}$CO ratio

The C abundance is expected to increase in the external layers of GMCs where the FUV radiation dissociates the CO, so that the [CI]/CO ratio may be greater than in the innermost parts of the cloud. To investigate the [CI]/$^{13}$CO in the ring we regrid the $^{13}$CO cube to the [C I] coarser resolution and applied to both the same [C I] 3D mask. In Fig. 21 we present the integrated intensity maps ratio between [C I] and $^{13}$CO.

The ratio between the zeroth moment maps possess values <1 along almost all the lines of sight passing through the thickest region of the molecular ring, with lower peaks in [NW-R] and [N-R]. Ratio values are >1 at the edge of the cloud subregions [SE-R], [SW-R], [C-R], and [N-R], although here $^{13}$CO detection drastically decreases, lowering the signal-to-noise level and making pixels less distinguishable. Interestingly, some sections of the ring edge have a ratio <1, in [NW-R], in eastern [C2] and around $(\ell, b) = (332.25°, 0.0°)$.

To better characterize the ratio values dispersion, we explore the ratio voxels distribution along the ring velocity range considering only those included in the 3D [C I] mask (see Fig. 22, 2D heatmap scatter plot between $^{13}$CO and [CI]/$^{13}$CO). The median of the ratio for the whole cloud is equal to 0.5, and its tendency is to decrease with the increase of $^{13}$CO emission, following a $y_0 + \frac{a}{x-x_0}$ law, with $a = 0.43 \pm 5.2 \times 10^{-3}$, $y_0 = 0.27 \pm 3.8 \times 10^{-3}$ and $x_0 = (8 \pm 1) \times 10^{-3}$. There are few ratio >1 voxels that come from low $^{13}$CO emission, the analysis of their position with respect to the molecular ring confirms that all of them are at the edge of the CO molecular ring.

## 9 C$^{18}$O CLUMPS CLASSIFICATION

Since C$^{18}$O emission is generally weak, due to the low isotopic abundance ratio compared to $^{12}$CO and $^{13}$CO, its detection is mostly likely to be optically thin. This makes the C$^{18}$O line a probe through the entire molecular clouds column, sensitive in particular to the highest column density regions, when $^{13}$CO (and even more so $^{12}$CO), is optically thick. In order to investigate the highest column density clumps within the molecular ring, we decomposed it applying the CLUMPFIND algorithm (Williams, de Geus & Blitz 1994) to the C$^{18}$O emission cube, binned and smoothed in the same way as in moment maps computation (see Section 5.1). We limited our analysis to the ring velocity range; however an expansion of this work, performed on 10° of the Galactic Plane binned C$^{18}$O datacube, is presented in Braiding et al. (2018). After a series of tests we selected the value of 0.35 K, equivalent to 3.5$\sigma$, for both input parameters *inc* and *low*. We set a threshold of 10 voxels on the minimum voxel number for a clump to be detected. We verified for false detection comparing C$^{18}$O and $^{13}$CO spectral lines passing through each clump centroid. To avoid false detection of a single clump as two distinct clumps we set a minimum centroid distance equal to seven voxels; all the clumps passed this verification step. The final detections are summarized in Table A1 with their physical characteristics.

For each clump we derived the virial mass following MacLaren, Richardson & Wolfendale (1988):

$$M_{vir} = \beta \Delta V^2 R_c, \quad (21)$$

where $\beta$ can be equal to 210, 190, or 126, respectively, for a constant, $1/r$ or $1/r^2$ spherical density profile, and $\Delta V$ is the clump velocity FWHM in km s$^{-1}$. $R_c$ is its effective radius in pc defined as

$$R_c = \sqrt{r_{eq}^2 - r_b^2}, \quad (22)$$

where $r_{eq} = \sqrt{A/\pi}$ and $A$ is the projected area derived from the extension of the zero moment map for each clump. Compared to the peak temperature of each clump ($T_p^{18}$), region detection is extended to low temperatures where the beam size is different from the normal beam FWHM, the $r_b$ factor takes into account this correction:

$$r_b = b \times 0.605 \sqrt{\ln\left(\frac{T_p^{18}}{T_{min}}\right)}, \quad (23)$$

where $b$ is the beam FWHM and $T_{min}$ is the lowest temperature at which the clump extents.

The resulting total virial mass is $M_{vir}^{tot} = 4.9 \times 10^4 \, M_\odot$. In order to compare it to the total mass derived under LTE assumption ($M_{LTE}^{20}$) we applied the method described in Section 6 including only voxels belonging to a clump related to the ring. According to the different $T_{ex}$ used, the total LTE clump mass traced by C$^{18}$O is 0.8 $\times 10^5$, 1.1 $\times 10^5$, and 1.5 $\times 10^5 \, M_\odot$, respectively, for $T_{ex} = 10$, 20, and 30 K. These values are similar to the LTE masses calculated for voxels included in the 3D smooth mask (Table 4). Comparing the total mass of the C$^{18}$O clumps and the total mass of the ring derived from $^{13}$CO, under the LTE assumption and taking the same $T_{ex}$ for both tracers, the fraction of gas in C$^{18}$O clumps is 59 and 76 per cent, respectively, for $T_{ex} = 10$ and 30 K. Changing *inc* and *low* parameters in CLUMPFIND, from 0.35 to 0.4 or 0.3 K, respectively, decreases or increases the LTE mass by $\sim$20 per cent for all $T_{ex}$, as a consequence of smaller or bigger clumps being identified by the algorithm.

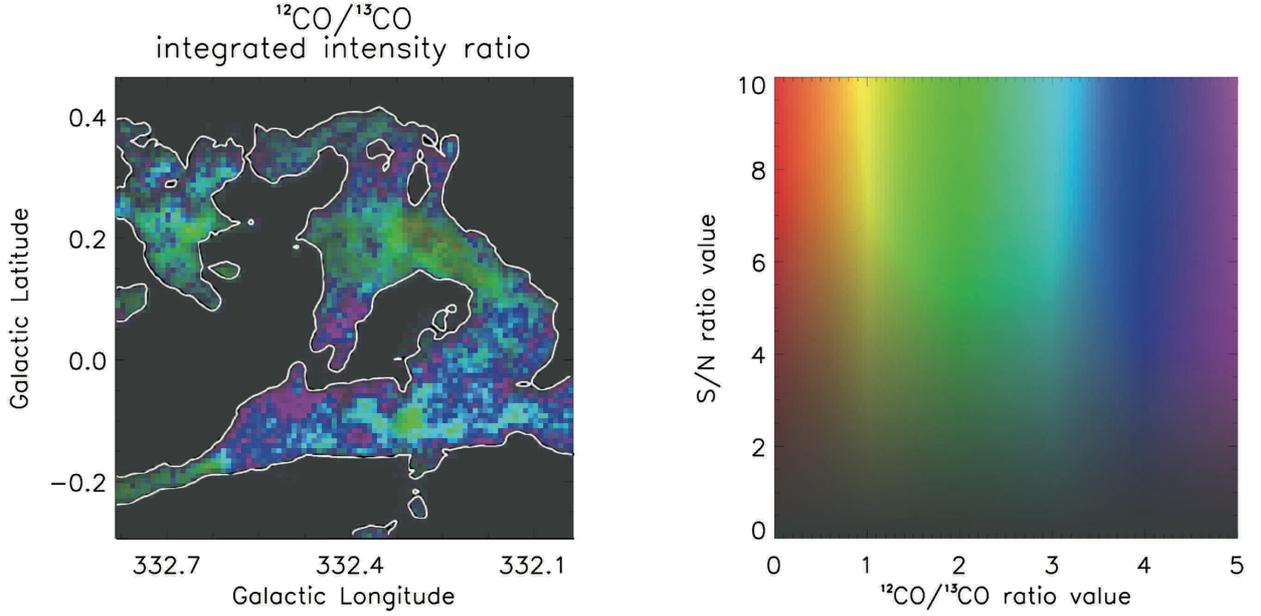

**Figure 20.** EVANS plot of the $^{12}$CO/$^{13}$CO integrated intensity map over the ring $V_{LSR}$ range. *Left-hand panel* – Each pixel in the ratio map possesses hue and saturation values that are decoded to SNR and ratio values using the colour table on the *right-hand panel* (see 8.1 for details on Evans plot). In white is the contour representing $^{13}$CO integrated emission at 3 K km s$^{-1}$.

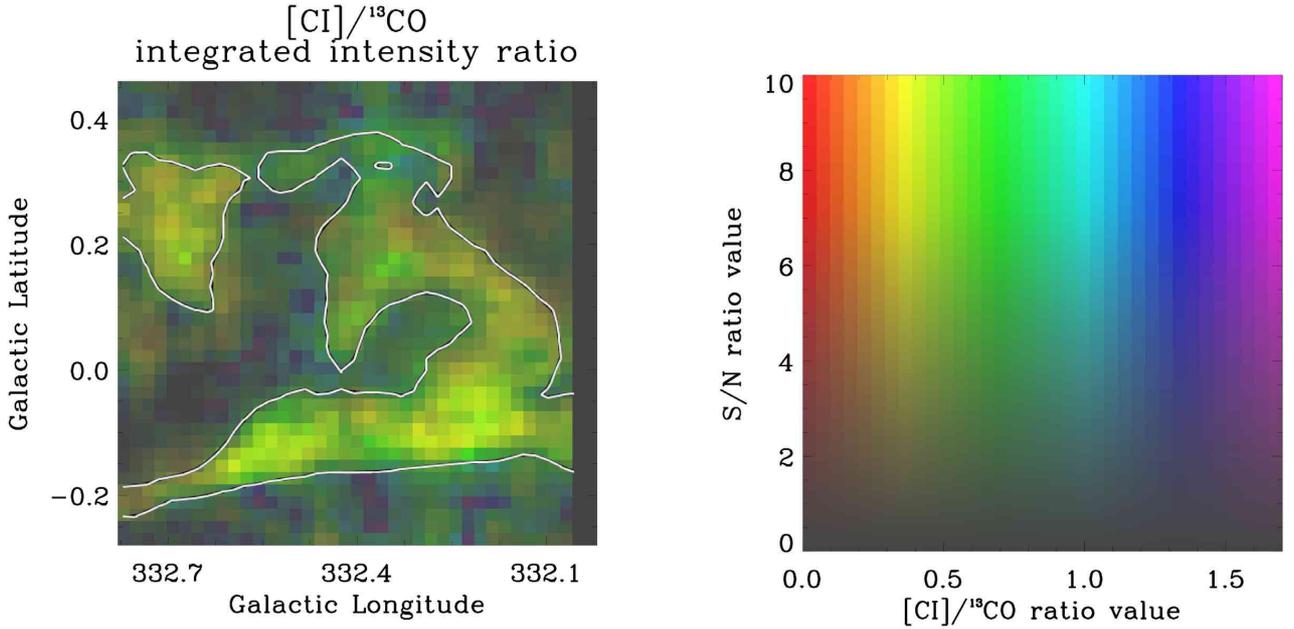

**Figure 21.** EVANS plot of the integrated intensity map [CI]/$^{13}$CO ratio along the ring $V_{LSR}$ velocity range. White contour is $^{13}$CO integrated intensity at 2 K km s$^{-1}$.

We note that the clumpfind algorithm found four clumps that are in a portion of the sky in which the C$^{18}$O integrated intensity has values equal to zero. This discrepancy is due to the C$^{18}$O 3D mask process creation. As described in Section 5.1, the 3D mask is derived from the smoothed cube. This operation decreases the intensity level of a weak emission extended for few voxels. As a consequence, this emission could result below the sigma threshold set on the smoothed cube in order to define the 3D mask extension. These *off mask* clumps are indicated with A, B, C, and D in Fig. 23 and, respectively, correspond to GMCC332.50+0.32, GMCC332.39+0.35, GMCC332.42+0.02, and GMCC332.29-0.01 in Table A1. Because of the exclusion from the 3D mask, the mass of these clumps is not present in the N(H$_2$) derivation as in Section 6 and reported in Table 4. Assuming a $T_{ex} = 20$ K their total mass results to be $10^4$ M$_\odot$, equal to ~8 per cent of the total C$^{18}$O clump mass.

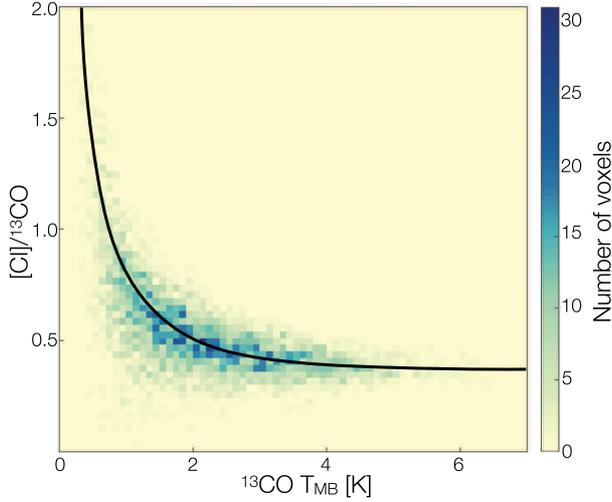

**Figure 22.** 2D heat map between $^{13}$CO datacube regridded to [C I] resolution and [CII]/$^{13}$CO voxels. The black line represents the fit in form: $y_0 + \frac{a}{x-x_0}$ (parameters are reported in Section 8.3).

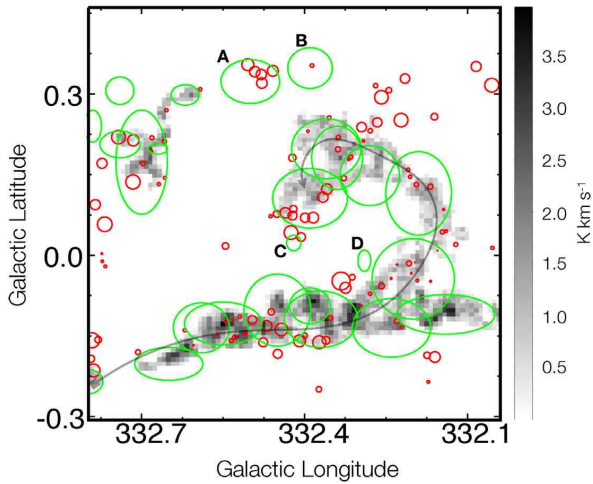

**Figure 23.** Plot of Hi-GAL compact sources (see Section 10.2) with a near distance solution between [3,4] kpc (red) and C$^{18}$O clump ellipsoids as reported in Table A1 (green). The black spline is the path along which are plotted the Hi-Gal dust clumps properties as in Fig. 27.

The virial parameter, defined as $\alpha_{vir} = M_{vir}/M_{LTE}^{20}$ and plotted in the Table A1 (column 11), gives an indication about the dynamical state of each clumps. Typical values are in the range [0.2–0.4] suggesting that they are not being supported against gravitational collapse. Though care is needed in this interpretation since masses derived under LTE assumptions applied depend on the $T_{ex}$ and C$^{18}$O abundance ratio choice.

## 10 DISCUSSION

In the following sections we will describe a method, based on the $^{13}$CO emission, to separate dust emission across the Galactic spiral arms, giving an estimate of the total mass traced by dust through the ring region. We will also discuss the different physical conditions in the cloud and how these suggest a scenario in which the molecular structure shows different evolutionary stages: quiescent clouds in which star formation is occurring at its earliest stage or not yet initiated, and active clouds in which star formation is ongoing at a later stage. In doing so, we will compare the C$^{18}$O with Hi-Gal clumps associated with the G332 ring cloud.

### 10.1 Dust across the Galaxy

As discussed in Section 6.5, the overall N(H$_2$) derived from the dust emission is the sum of all contributions along the line of sight. To compare it to N(H$_2$) traced by the $^{13}$CO we first consider Fig. 14, relating dust column to $^{13}$CO line flux. There are two sets of point here, one (group A) showing a linear correlation, the other (B) not. The former demonstrates the link between flux and column density for most regions, which we now decompose into contributions coming from separate spiral arms along the line of sight.

We note that in the N(H$_2$) column density traced by dust (Fig. 13, which comprises all infrared emission from this region) it is possible to find, to first approximation, a coherency between the dust and $^{13}$CO and $^{12}$CO integrated flux (Fig. 4). In particular, as shown in Fig. 13, in [SE-R] and [SW-R] $^{13}$CO integrated flux peaks are close to dust column peaks. The West side of the ring, [W-R], and part of [NW-R], present low integrated CO emission while the dust is bright. Bright dust, in general, follows the ring, a likely consequence of it arising from the region too. However, focusing on the ring CO and dust emission, we need to separate dust related to different Galactic spiral arms. In the following we describe our approach.

According to the average CO spectral line profile and following the spiral arm scheme adopted in this paper (Fig. 5), we separated the $^{13}$CO emission into four velocity ranges (R1, R2, R3, and R4), choosing the near or far solution coherently with any H I emission local minimum (see Fig. 3):

(i) R1: [$-115$ to $-75$ km s$^{-1}$] [intra-arm Perseus/Norma far]
(ii) R2: [$-74$ to $-56$ km s$^{-1}$] [Norma far]
(iii) R3: [$-55$ to $-44$ km s$^{-1}$] [Scutum-Crux near] (ring)
(iv) R4: [$-44$ to $-30$ km s$^{-1}$] [Scutum-Crux near]

This scheme follows the position velocity (PV) plot of Fig. 6. We excluded $^{13}$CO emission coming from $-30$ to $+10$ km s$^{-1}$ since it is localized in small structures and its integrated intensity is negligible with respect to other velocity intervals. To separate the dust emission coming from each of the spiral arms we computed, following the method of Section 5.2, for each velocity interval, a $^{13}$CO pixel-by-pixel moment zero map normalized to the total integrated intensity relative to the $V_{LSR}$ interval [$-115, -30$ km s$^{-1}$] (see Fig. 24), each of which is subsequently multiplied by the N(H$_2$) derived from dust SED fit (Fig. 13). In this way we created four different N(H$_2$) dust maps (see Fig. 25 one for each spiral arm). Our approach is based on two assumptions that N(H$_2$) traced by $^{13}$CO is proportional to $^{13}$CO integrated intensity and that the emission from dust coming from regions where there is no CO emission is negligible (see below for H I contributions). Thus, we are able to compare a directly measured quantity ($^{13}$CO emission) to N(H$_2$) dust. We also excluded the molecular hydrogen not traced by CO in this analysis.

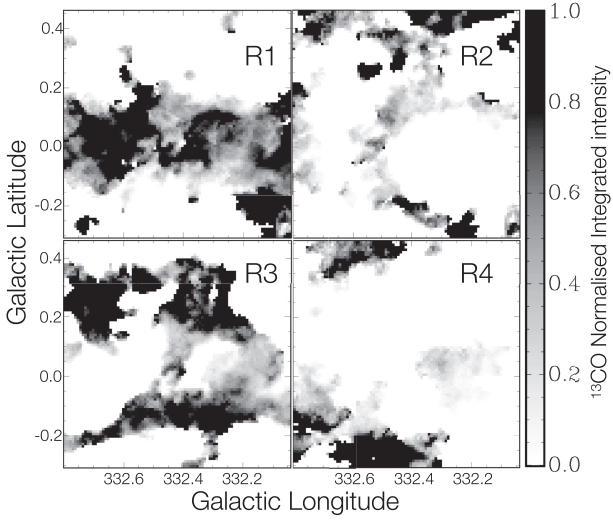

**Figure 24.** Normalized $^{13}$CO integrated intensity map belonging to the defined velocity ranges as in Section 10.1. Each pixel represents the multiplication factor applied to the corresponding pixel in the dust derived N(H$_2$) map; thus it is possible to obtain a fractional molecular hydrogen column density traced by dust for each velocity range.

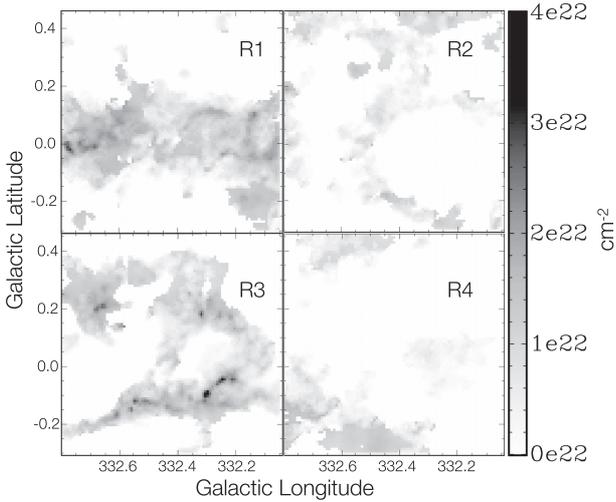

**Figure 25.** N(H$_2$) dust maps relative to each velocity interval derived from the $^{13}$CO normalized moment zero maps plotted in Fig. 24 multiplied by the N(H$_2$) dust map derived from the dust SED fit (Fig. 13).

Scatter plots, between $^{13}$CO integrated intensity and N(H$_2$) dust for each spiral arm, are shown in Fig. 26. Higher column densities belong to [R1] and [R3]. In each plot we use different colours for those pixels having a normalized $^{13}$CO integrated intensity map equal to 1, while the other pixels are in grey. Both of them follow a direct proportional relation but have not the same offset: grey points starts at zero while coloured pixels begins at about $10^{22}$ cm$^{-2}$ for all the ranges. Pixels belonging to the *offset* group comes from regions in which the normalized $^{13}$CO moment zero map has values equal to 1, which means the total $^{13}$CO integrated intensity (along the whole line of sight) has contributions coming from a single velocity interval only, without contamination from other clouds at different distances. These are those pixels in black (value equal to 1) in the normalized moment zero maps (Fig. 24); for them dust emission can be constrained in a specific velocity range since there is no other CO detection on the same line of sight.

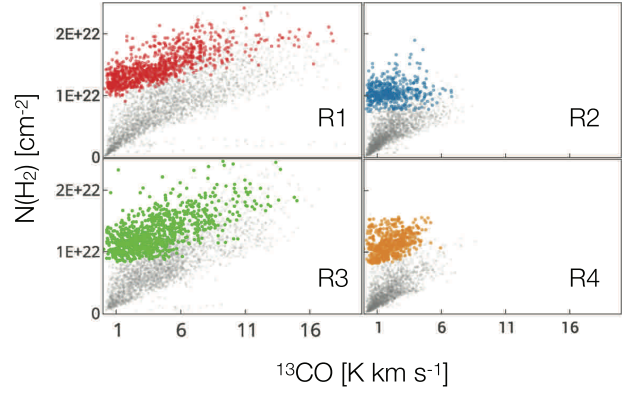

**Figure 26.** Scatter plots between normalized moment zero $^{13}$CO ($x$ axis) and fractional N(H$_2$) dust relative to all velocity intervals (as defined in Section 10.1). For each scatter plot we use colours for those pixels whose normalized $^{13}$CO integrated intensity map is equal to 1 (see Fig. 24). The grey points represent all the other pixels.

**Table 6.** Linear fit parameters for each velocity interval from plots in Fig. 26 derived using the unity value pixels in the normalized moment zero maps (coloured pixels). Numbers between parenthesis are interval widths.

| | Fit parameters | |
|---|---|---|
| | N(H$_2$)$_{dust}$ = m · F$_{13CO}$ + b | |
| Interval | m | b |
| (km s$^{-1}$) | (cm$^{-2}$K$^{-1}$ km$^{-1}$ s) | (cm$^{-2}$) |
| R1 (40) | $6.1 \times 10^{20} \pm 0.2 \times 10^{20}$ | $1.21 \times 10^{22} \pm 1.0 \times 10^{20}$ |
| R2 (18) | $1.3 \times 10^{20} \pm 0.6 \times 10^{20}$ | $1.03 \times 10^{22} \pm 1.3 \times 10^{20}$ |
| R3 (11) | $7.4 \times 10^{20} \pm 0.3 \times 10^{20}$ | $1.01 \times 10^{22} \pm 1.5 \times 10^{20}$ |
| R4 (14) | $6.1 \times 10^{20} \pm 0.7 \times 10^{20}$ | $1.02 \times 10^{22} \pm 1.5 \times 10^{20}$ |

In Table 6 are reported linear fit parameters for each velocity interval, considering only coloured points as in Fig. 26. The offset value is almost the same for every range, around $10^{22}$ cm$^{-2}$, this value is similar to the mean H$_2$ column density traced by dust in regions where no CO emission is detected along the whole Galactic velocity range (considering the $^{12}$CO integrated intensity map in the [$-110, +10$ km s$^{-1}$] $V_{LSR}$ range). This could be a consequence of the infrared background emission coming from this part of the sky, [R1] shows the largest background having the largest velocity width. The fit parameters (except for [R2], but this has the lowest contribution to the total emission), are comparable to the fit parameters reported in Fig. 14. The relative contributions of each velocity range with respect to total $^{13}$CO integrated intensity, along the whole line of sight, are 35, 8, 47, and 10 per cent, respectively, for R1, R2, R3, and R4. Although, R3 is only 10 km s$^{-1}$ wide, it accounts for almost half of the total $^{13}$CO emission in G332.

Taking into account a near or a far distance solution, the mass traced by the disentangled dust in each velocity interval is reported in Table 7, together with the mean neutral hydrogen column density, N(HI), which is the 13, 13, 6, and 8 per cent of N(H$_2$) column density (derived from dust) in [R1], [R2], [R3], and [R4], respectively.

Comparing HI and CO column densities, across the $-110 < V_{LSR} < +10$ km s$^{-1}$ velocity range, we found that the amount of atomic gas is about ∼10 per cent of the column density of the molecular gas, so that the dust column density is dominated by emission from dust within molecular clouds. The total mass traced by dust across the whole sky portion defined by the ring region in the Galaxy velocity range ($-110 < V_{LSR} < +10$ km s$^{-1}$) is ∼$2.15 \times 10^6$ M$_\odot$.

**Table 7.** Summary table of the four velocity ranges in which is divided the dust emission of the G332 Galactic plane sector. The mean N(H$_2$) is traced from the fractional dust emission as described in the Section 10.1. The distance is computed from the $V_{LSR}$ value of the centroid velocity of each interval.

| Interval | Distance kpc | Mass M$_\odot$ | Mean N(HI) $10^{20}$ cm$^{-2}$ | Mean N(H$_2$) $10^{20}$ cm$^{-2}$ |
|---|---|---|---|---|
| R1 | 8.8 | $12.2 \times 10^5$ | 13.0 | 100 |
| R2 | 10.5 | $6.6 \times 10^5$ | 6.9 | 54 |
| R3 | 3.7 | $2.00 \times 10^5$ | 6.1 | 95 |
| R4 | 2.9 | $0.5 \times 10^5$ | 4.6 | 56 |

**10.2 Star formation activity along the molecular ring**

The overview image in Fig. 1 shows the Southern side of the ring hosting several signposts of star formation with activity (masers and H II regions) counterpoised a less active Northern side. In order to investigate this difference we compared the C$^{18}$O clumps distribution (reported in Table A1) to the compact sources distribution taken from the Hi-GAL catalogue (Elia et al. 2017) classified for different stages of star evolution.

The Hi-GAL sources (hereafter called dust clumps) selected for our analysis have coordinates that fall inside the C$^{18}$O clumps ellipsoid extension (with semi-axis defined by $\Delta \ell$ and $\Delta b$ FWHMs derived by clumpfind), and a near distance solution between 3 and 4 kpc. In order to avoid the exclusion of sources not satisfying these criterion, but potentially linked to the molecular ring, we included all the dust clumps that totally (or partially) fall in the portion of sky covered by the angular extension of the C$^{18}$O integrated intensity map (see Fig. 23). The total mass of all the dust clumps considered is ≈$6.8 \times 10^4$ M$_\odot$, equal to the 35 per cent of the ring mass traced by $^{13}$CO at $T_{ex} = 30$ K.

Following Veneziani et al. (2017), we quantified the *Clump Formation Efficiency* (CFE), defined as

$$CFE = \frac{M_{clumps}}{M_{GMC}}. \quad (24)$$

In this way it is possible to infer the molecular cloud mass fraction that could be involved in the star formation process, in cores in which star formation could occur. We used as M$_{clumps}$ both *pre-* and *proto-stellar* dust clump masses, while M$_{GMC}$ is the molecular cloud mass traced by C$^{18}$O for different assumed $T_{ex}= 10$, 20, and 30 K, $T_{map}$ and $T_{dust}$ (as in Table 4) that, respectively, gives a CFE= 72, 53, 41, 51, and 58 per cent.

The L$_{bol}$/M$_{clump}$ ratio is a useful indicator of the star formation activity ongoing in the hosting clump (Molinari et al. 2008; Elia & Pezzuto 2016). In order to examine the star formation activity along the molecular ring, in Fig. 27 we plot the L$_{bol}$/M$_{clump}$ and T$_{clump}$ values of the dust clumps following the path defined from the [E,SE-R] to the northern part of [C-R] (see Fig. 23).

The highest dust clump temperatures are in the Southern ring, as well as the highest L$_{bol}$/M$_{clump}$ ratios, both decrease following the path. The C$^{18}$O clumps corresponding to the Southern ring (as reported in Table A1) exhibit in mean, higher values in $^{13}$CO, C$^{18}$O average temperatures and in velocity dispersion, while those relative to the Northern cloud show lower values for the same quantities. These factors suggest a difference in the star formation activity along the cloud: the dust cores in the Southern ring exhibits (on average) hotter temperatures suggesting that star formation activity is going on, while the Northern cores, colder, indicate that the star formation has not yet started, or at an earlier stage.

**10.3 Local X$_{co}$ variation**

$^{12}$CO ($J = 1 \rightarrow 0$) emission is often used in the determination of the N(H$_2$), and total molecular mass, through the use of the X$_{co}$ factor, especially for extra-Galactic sources. Thus, it is important to understand how the physical conditions of the emitting gas contribute to any variation in the X factor. Considering the cloud distance and the pixel dimension of 30 arcsec we are able to examine X$_{co}$ variations on parsec scales (at ∼0.5 pc).

The approach of our X$_{co}$ analysis uses equation (7), where N(H$_2$) is traced by $^{13}$CO under different T$_{ex}$ assumptions, or N(H$_2$) traced by dust. As shown in Section 7.1, the average X$_{co}$ is found to increase with the assumed T$_{ex}$ used to derive the column density.

A pixel-by-pixel map of the cloud X$_{co}$ values is shown in Fig. 18, the highest values are in [NW-R], and in part of [N-R] (at $b \sim 0.2°$ and $332.45° < \ell < 332.15°$), in [A6], in [B6], and around the position ($\ell$, $b$) = ($332.3°$, $-0.1°$) of typically ∼$5 \times 10^{20}$ cm$^{-2}$ K$^{-1}$ km$^{-1}$ s.

$^{12}$CO spectral lines, where the X$_{co}$ is higher, show lower $^{12}$CO intensity compared to ring average, as does $^{13}$CO as well, and in general present a non-Gaussian spectrum profile. In many cases, $^{12}$CO is self-absorbed, while $^{13}$CO and C$^{18}$O peak at the $^{12}$CO local minimum (as shown in Fig. 7).

Local decrease in intensity of the $^{12}$CO/$^{13}$CO ratio (Fig. 20) corresponds to higher X$_{co}$ factor (Fig. 18), where the $^{12}$CO/$^{13}$CO ratio is lowest, the CO is most optically thick and the X$_{co}$ is highest.

This is not unexpected since, by definition, our X$_{co}$ map derives from the ratio between N(H$_2$) traced by $^{13}$CO (proportional to $^{13}$CO integrated intensity through the optical depth integration; see equations 4 and 5) and $^{12}$CO integrated intensity. These optically thickest regions show significant decrease in $^{12}$CO emission with respect to $^{13}$CO.

*10.3.1 X$_{co}$ and T$_{ex}$*

To further investigate the physical conditions underlying the observed X$_{co}$ distribution (Fig. 19), we looked at the variation of T$_{ex}$ around the molecular ring. In Fig. 28 are shown the scatter plots of T$_{ex}$ versus X$_{co}$ for each sub-region of the cloud. The T$_{ex}$ value was derived from the $^{12}$CO peak temperature for each sight line, while the X$_{co}$ was computed by using the N(H$_2$) traced by $^{13}$CO, assuming in the $^{13}$CO column density a T$_{ex}$ derived from the dust temperature computed by the SED fit (see Section 6.5). In Table 8 are summarized the X$_{co}$ and T$_{ex}$ mean values for each region.

Examining Fig. 28 we note a general decrease in the dispersion of X$_{co}$ values, moving from [N-R] towards [SE-R], while T$_{ex}$ increases. This behaviour in T$_{ex}$ is similar to the temperatures and L$_{bol}$/M$_{clump}$ ratio trend observed in the dust clumps (Fig. 23) and is consistent with a stellar evolution gradient scenario: the Southern cloud hosts active and more evolved star formation sites, which heat the surrounding gas so increasing its temperature. On the contrary, the Northern cloud hosts less and, on average, earlier stage star formation sites, the gas is colder compared to the South and the peak

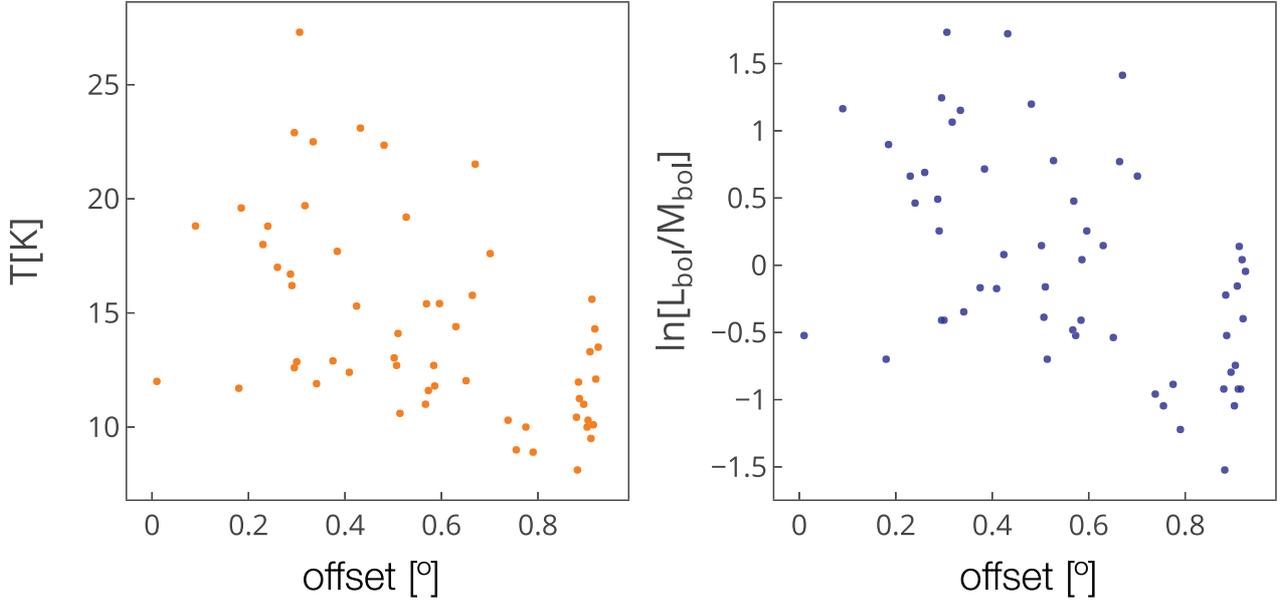

**Figure 27.** Plot of Hi-Gal $T_{clump}$ (left-hand panel) and of $\ln[L_{bol}/M_{clump}]$ (right-hand panel) ratio derived from the bolometric luminosity of the clump and its mass (in $M_\odot$). The *x* axis is the offset distance from the initial point of the path (at coordinates $\ell \sim 332.75°$ and $b \sim -0.25°$) defined in Fig. 23 by the black spline.

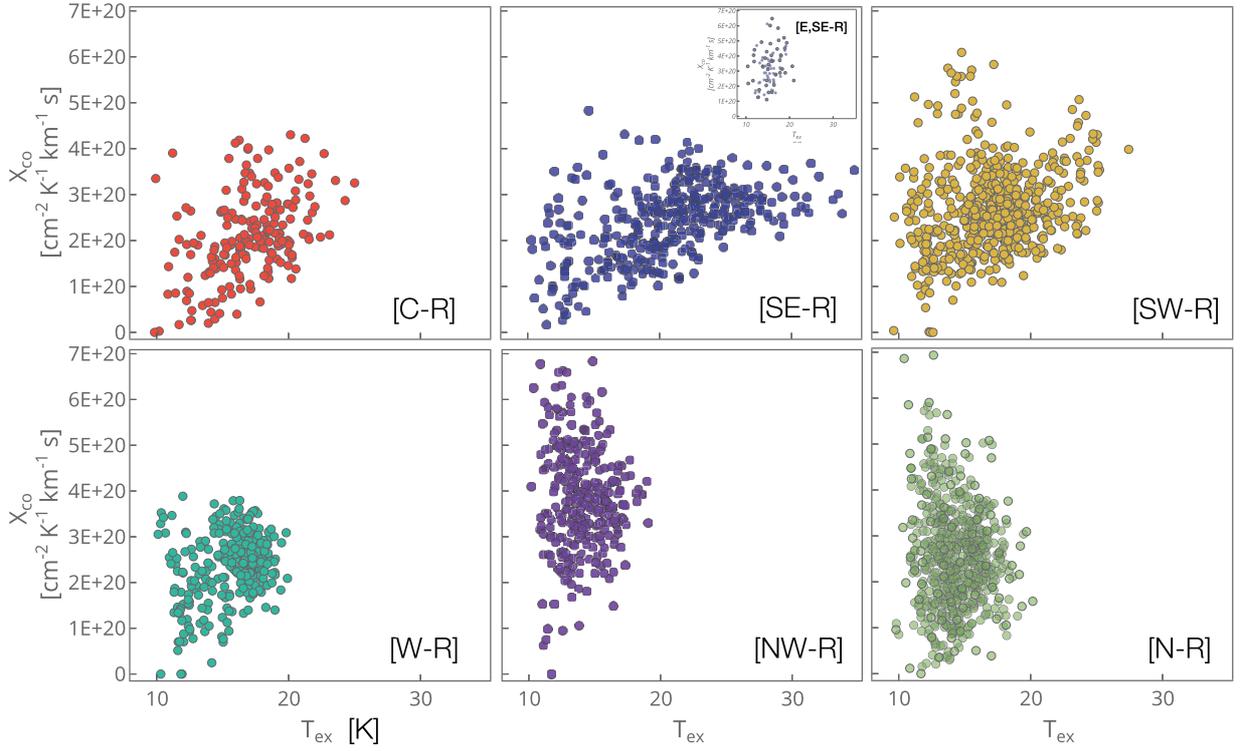

**Figure 28.** $X_{co}$ factor plots for each ring sub-region as per Fig. 1. The range of *x* and *y* axes are the same for each plot. The plot [SE-R] comprises only pixels coming from [SE-R], excluding those with $\ell > 332.65°$, which are reported in the sub-plot indicated with [E,SE-R] (*Eastern-South East, Region*).

**Table 8.** Summary table of mean $X_{co}$, $T_{ex}$, and mean [CI]/$^{13}$CO ratio values per region in Fig. 28. The mean value for the $T_{ex}$ is derived from the mean peak emission of the $^{12}$CO along the line of sight of each region.

| Region | Mean $T_{ex}$ K | Mean $X_{co}$ cm$^{-2}$ K$^{-1}$ km$^{-1}$ s | [CI]/$^{13}$CO |
|---|---|---|---|
| [N-R] | 14 | $2.6 \times 10^{20}$ | 0.7 |
| [NW-R] | 15 | $3.8 \times 10^{20}$ | 0.5 |
| [W-R] | 16 | $2.5 \times 10^{20}$ | 0.5 |
| [SW-R] | 18 | $2.6 \times 10^{20}$ | 0.6 |
| [SE-R] | 21 | $2.4 \times 10^{20}$ | 0.6 |
| [E, SE-R] | 15 | $3.3 \times 10^{20}$ | 0.4 |
| [C-R] | 17 | $2.1 \times 10^{20}$ | 0.7 |

emissions are the lowest, as well as the velocity dispersion and the column densities.

Supporting a scenario in which the Southern cloud is more active than the Northern, as illustrated in Fig. 1, the literature reports many indicators of star formation activity, as H II regions (see Section 5.2) and masers, localized in the Southern ring: 6 GHz methanol maser (Caswell et al. 2010), water maser (Walsh et al. 2014), class I methanol maser at 36 and 49 GHz (Voronkov et al. 2014), and 95 GHz maser (Yang et al. 2017).

Recently Kong et al. (2015) reported for the California Molecular Cloud high $X_{co}$ values dispersion for cold quiescent regions and an inverse correlation between $X_{co}$ and $T_{ex}$. They argued that this correlation could explain the range of $X_{co}$ values, found for different GMCs, as a consequence of the different amount of warm gas heated by OB stars in the cloud. The G332 molecular cloud presents similar behaviour, where the coldest parts of the cloud show the most scattering and highest $X_{co}$ values and the lowest $T_{ex}$, while the most active regions show the opposite behaviour.

## 10.4  [C I]

The UV radiation coming from the massive young stars photodissociates the surrounding gas, creating a PhotoDissociation Region (PDR), where the carbon is mainly in atomic form, in neutral or singly ionized state (Tielens & Hollenbach 1985; van Dishoeck & Black 1988); similarly, the external layer of molecular clouds are exposed to the diffuse far-UV radiation, photodissociating the CO before it reaches a column density high enough to enable self-shielding. To inspect the atomic carbon distribution around the cloud, in Fig. 21 is illustrated the ratio between the integrated intensities of [C I] and $^{13}$CO. As expected, the outskirts of the clouds possess the highest ratio values, though with the lowest SNR. Interestingly, [NW-R] and [W-R] present an exception: low ratio values at the edge facing the North–West direction, while the edge towards South–East presents higher values. Looking at the sightlines passing through the innermost part of the cloud, the regions with the lowest [CI]/$^{13}$CO ratio values are [N-R], [NW-R], [E,SE-R], and part of [W-R]. Profiles showing the ratio as a function of velocity are in Fig. 29. They were derived from the [C I] and $^{13}$CO (regridded to the [C I] resolution) datacubes, both masked using the same [C I] 3D mask, the resulting profiles were then plotted applying a Hanning smooth of order 3. The resulting picture is of a northern ring, [N-R] and [NW-R], with mean ratio values of 0.3 centred at $V_{LSR} = -50$ km s$^{-1}$ (we consider the second peak at $V_{LSR} = -42$ km s$^{-1}$ belonging to another cloud; see Fig. 3), while the southern ring has a mean ratio value of 0.5, peaking at 0.6 in the external layer of the cloud. In [SW-R] (aperture D) the mean ratio spectral profile shows a constant value going deeper in the cloud, a possible indication of photodissociation in the cloud interior caused by star formation activity. Apertures A and C, belonging to the southern cloud, show a small dip in the innermost part of the ring, where the $^{13}$CO emission increases. Regarding aperture E, we note that the emission velocity centroids of the [C I] and the $^{13}$CO emissions are close to $V_{LSR} = -50$ km s$^{-1}$ (Fig. 8), while we consider the second peak at $V_{LSR} = -42$ km s$^{-1}$ belonging to another cloud.

We conclude that the fraction of atomic carbon, with respect to its molecular form, is lower in the Northern cloud than in the Southern cloud. The latter possesses the highest photodissociation levels due to increased radiation fields. We speculate that [NW-R] is the most quiescent part of the cloud, and the difference between North and South is driven by a star formation activity ongoing at a later stage in [SW-R] and [SE-R], compared to [N-R], [NW-R], and [E,SE-R]. This further strengthens the hypothesis of a difference in the star formation activity between northern and southern ring.

## 10.5  CO and 8 μm emission

IRDCs are regions in which the gas is cold and dense, absorbing background IR emission. They provide suitable targets to look for signposts of early star formation activity (Teyssier, Hennebelle & Pérault 2002; Rathborne, Jackson & Simon 2006; Simon et al. 2006). The dust temperature map reported in Fig. 11 suggests that the North–Western ring is the coldest region, thus it is likely to host IRDCs. In Fig. 30 we plot the angular positions and equivalent radius size of the IRDCs (red circles) identified by Peretto et al. (2016), overlaid on the $X_{co}$ map. Despite the coarser angular resolution of the $X_{co}$, compared to the 16 arcsec Hi-GAL maps, it is possible to see a spatial correlation between their positions and the angular extent of the molecular cloud.

In Fig. 31 is shown the Spitzer 8μm image, in green contour is the angular extension of the $^{13}$CO integrated intensity map, while the red contours represents the dust temperature as per Fig. 11. We note that the distribution of cores with lowest dust temperatures precisely follow the extinction features seen in 8μm. Moreover, between $332.3° < \ell < 332.1°$ and $-0.1° < b < 0.3°$, an arc-like extinction feature is clearly seen, following the CO emission in the sky portion defined by northern [SW-R], [W-R], and [NW-R] and southern [N-R]. The same correspondence between lowest dust temperatures values and 8μm extinction is seen in [E,SE-R]. On the other hand, the highest dust temperatures match the highest 8μm emission. We found correspondence between the distribution of N(H$_2$) traced by C$^{18}$O and the extinction features (see Fig. 32), also in [E,SE-R], and in small localized regions of the southern cloud, i.e. $(\ell, b) = (332.4°, -0.1°)$, suggesting a strong correlation between the IRDC position, the gas in the molecular cloud, and the coldest dust.

Carbon monoxide depletion has been observed (Alves, Lada & Lada 1999; Kramer et al. 1999; Kong et al. 2015) in dense and cold regions. Thus, the cold and dense environment of IRDCs could favour CO depletion caused by CO removal from gas phase due to freeze out on to dust grains. It is important to note that could affect the CO to dust comparison method presented in Section 10.1, which was used to decompose the dust emission along the line of sight across the Galaxy spiral arms, since we assumed that the dust and CO emission are proportional. A consequence of the CO depletion would be an underestimate of the dust content.

Comparing N(H$_2$) traced by dust and N(H$_2$) traced by C$^{18}$O to probe the densest regions of molecular cloud, we found the distribution of N(H$_2$) by C$^{18}$O follows the IRDC position in the Northern part of [SW-R], in [W-R] and [NW-R], as well as in localized regions of the southern ring. In particular, in cold dust regions (Fig. 33), we found an offset between the $C^{18}O$ N(H$_2$) peaks and dust N(H$_2$) peaks, which could be an indication of CO depletion in the highest column density dust regions.

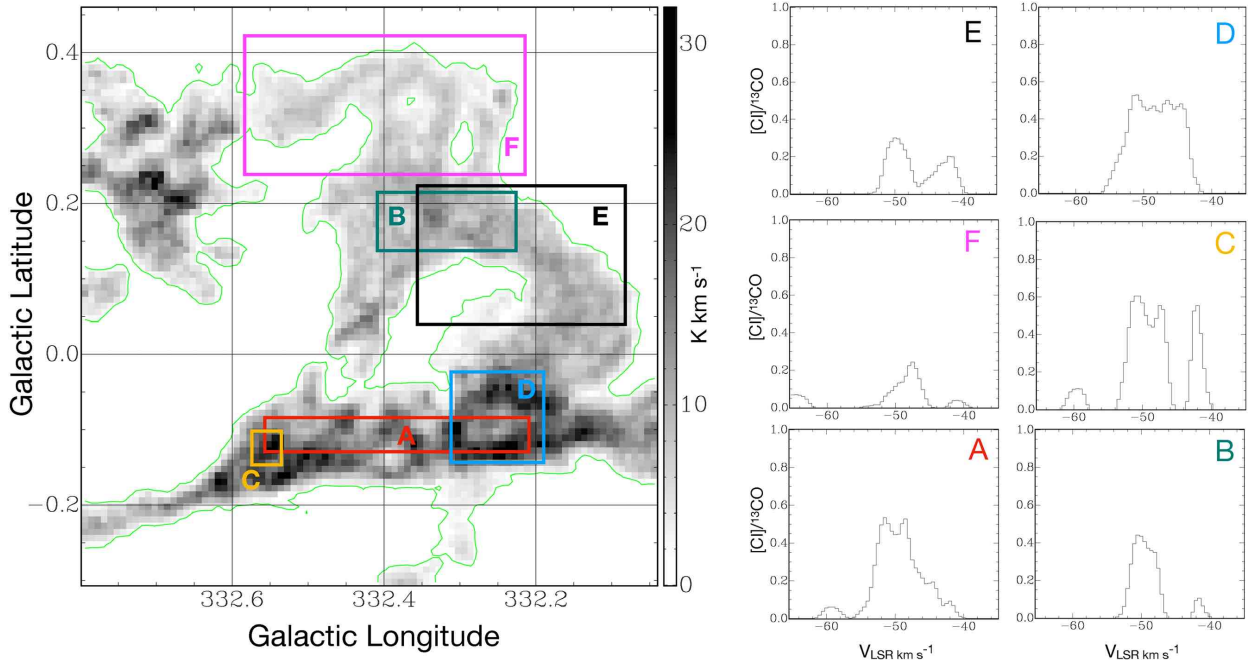

**Figure 29.** Plots of the mean spectral profiles of [CI]/$^{13}$CO ratio *right* for the different apertures indicated with different letters in the *left-hand image*. The image map represents the $^{13}$CO integrated intensity map, with the green contours its extent.

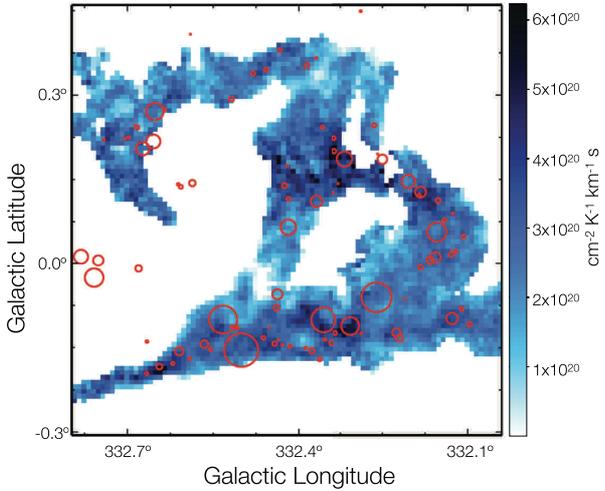

**Figure 30.** Image of the $X_{co}$ factor (see Section 7.1) calculated assuming as $T_{ex}$ the $^{12}$CO peak temperature map. The red circles represent the angular position and the equivalent radius of the IRDCs (as reported by Peretto et al. 2016) .

The emission at 8 $\mu$m comes mainly from PAHs (Tielens & Hollenbach 1985) excited by UV radiation from high-mass stars in PDRs. The highest 8$\mu$m emission (in Fig. 31) are localized in [SW-R], [SE-R], and [C-R] regions, where the CO spectral lines show the highest emission peaks and the highest velocity dispersion. If the gas in the molecular cloud and the IRDC are part of the same structure, we expect to observe a decreasing gradient in the $T_{ex}$ from [SE-R] to [N-R], similar to that observed from the Southern to Northern cloud (see Table 8).

Taking into account all these factors, we hypothesize that this IRDC arc-like structure, and its extension in the souther ring, is associated with the coldest and densest parts of the cloud, a unique structure along which the star formation activity is at a later stage in the southern ring than the northern ring. Considering [E,SE-R] and [N-R] as the extremities of the IRDC, its length is ∼65 pc.

## 11 VIRTUAL REALITY AND AUGMENTED REALITY: AN APPLICATION TO RADIO ASTRONOMY

In this section we present a new perspective for looking at astronomical data, developed during the analysis of the G332 molecular ring and applicable to other 3D surveys. The aim is to find a visualization method to better understand the distribution of CO data and to perform an efficient multiwavelength synergy with other surveys. This is not only a data-driven problem but also a data visualization issue, which we tried to solve by experimenting with the use of VR and AR environments and applications.

Currently, the astronomy community is experiencing a growing interest in AR/VR software and hardware to improve data visualization and analysis (Ferrand, English & Irani 2016; Fluke & Barnes 2018). However, most of the software used by astronomers does not offer advanced visualization tools, mainly because it was developed before the current available hardware, but also because its 3D plot capabilities have limited user interaction. On the other hand, the possibility given by VR/AR tools to literally fly through data sets dramatically improves the analysis process. Since our senses evolved in a 3D world, a more intuitive data interpretation is likely to occur in a 3D environment rather than in a 2D representation; this is why immersive techniques have the potential to boost data examination. We made use of VR/AR visualization in our data CO investigation[10] to discriminate between emissions related to the main molecular ring structure (as in Section 5.3) and not. Standard 2D datacube plots (as moments maps or PV plots) could lead to data misinterpretations, because of projection effects, while a 3D immersive visualization could help showing data in a greater dimensional space. To solve the above ambiguity, we inspected $^{12}$CO, $^{13}$CO, and C$^{18}$O emissions at different thresholds at the same time ($2\sigma$, $3\sigma$, $3.5\sigma$, $4\sigma$, and $4.5\sigma$). Such an amount of details is hard to clearly represent on a standard 3D plot, restricting analysis to a limited amount of information.

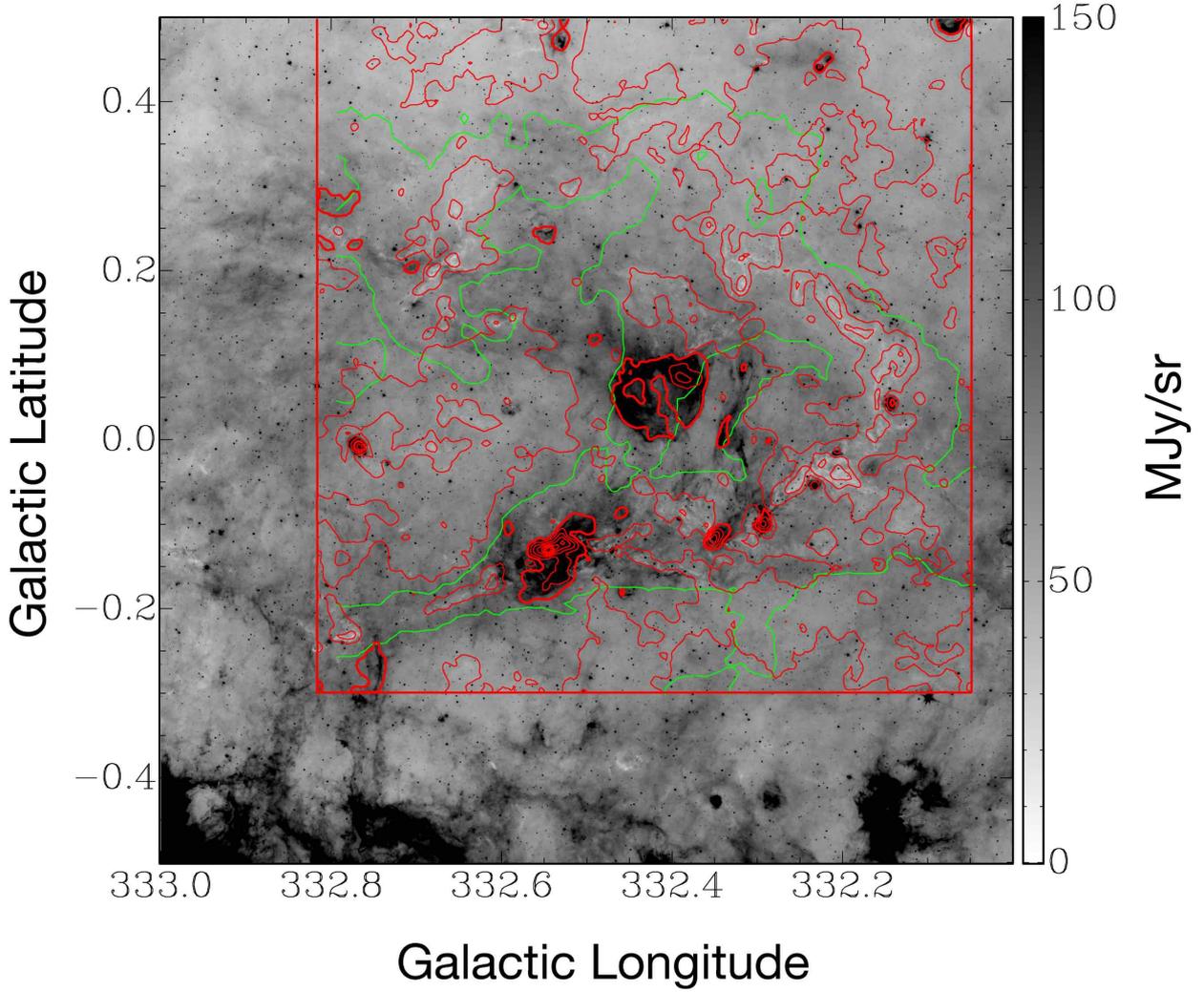

**Figure 31.** Spitzer 8 μm map, overlaid contours in green are the extent of the $^{13}$CO while in red is the dust temperature as per Fig. 11. The temperature contours are in the range [15,27 K] in steps of 1 K. The red square delimits where the CO data are used.

However, it is important to say that care is needed in PPV 3D datacube inspection since the velocity (or spectral) dimension could lead to misinterpretation in the analysis of the emission morphology (as discussed in Clarke et al. 2018).

A digression on the limits of VR/AR application in astronomy is beyond the purpose of this paper; they were analysed by Vogt & Shingles (2013). Here we note that the continuous increase of computing power, and the consequent increase in smartphone performance, makes it possible to use complex software capable of achieving advanced results without increasing its complexity.

In Appendix A we describe in detail three examples of our VR and AR visualization techniques. All of them are based on the same initial step: convert the FITS format into another 3D format, which can be easily read by VR (or AR) frameworks.

## 12 SUMMARY

We used data from the Mopra Southern Galactic Plane CO Survey of the $^{12}$CO, $^{13}$CO, and C$^{18}$O $J = 1 \rightarrow 0$ transition lines (Braiding et al. 2018) to perform a morphological and a physical characterization of a GMC (*G332 molecular ring*) sited in the G332 region of the Galactic plane. The CO emission peaks at $V_{\rm LSR} = -50\,\rm km\,s^{-1}$, which is equal to a distance of $\sim 3.3$ kpc from the Sun (assuming correct a near distance solution for KDA) and to a Galactocentric distance of $\sim 5.5$ kpc. The cloud shape is similar to a ring of $\sim 42$ pc in diameter. The total mass of the cloud is estimated in $\sim 2 \times 10^5\,M_\odot$, assuming [$^{12}$CO]/[$^{13}$CO] = 53, [H$_2$]/[$^{12}$CO] = $1.1 \times 10^4$, and uniform $T_{\rm ex} = 30$ K for all the subregions of the molecular structure. We found a strong spatial correlation between the C$^{18}$O and an IRDC structure that was seen as an extinction feature at 8 μm.

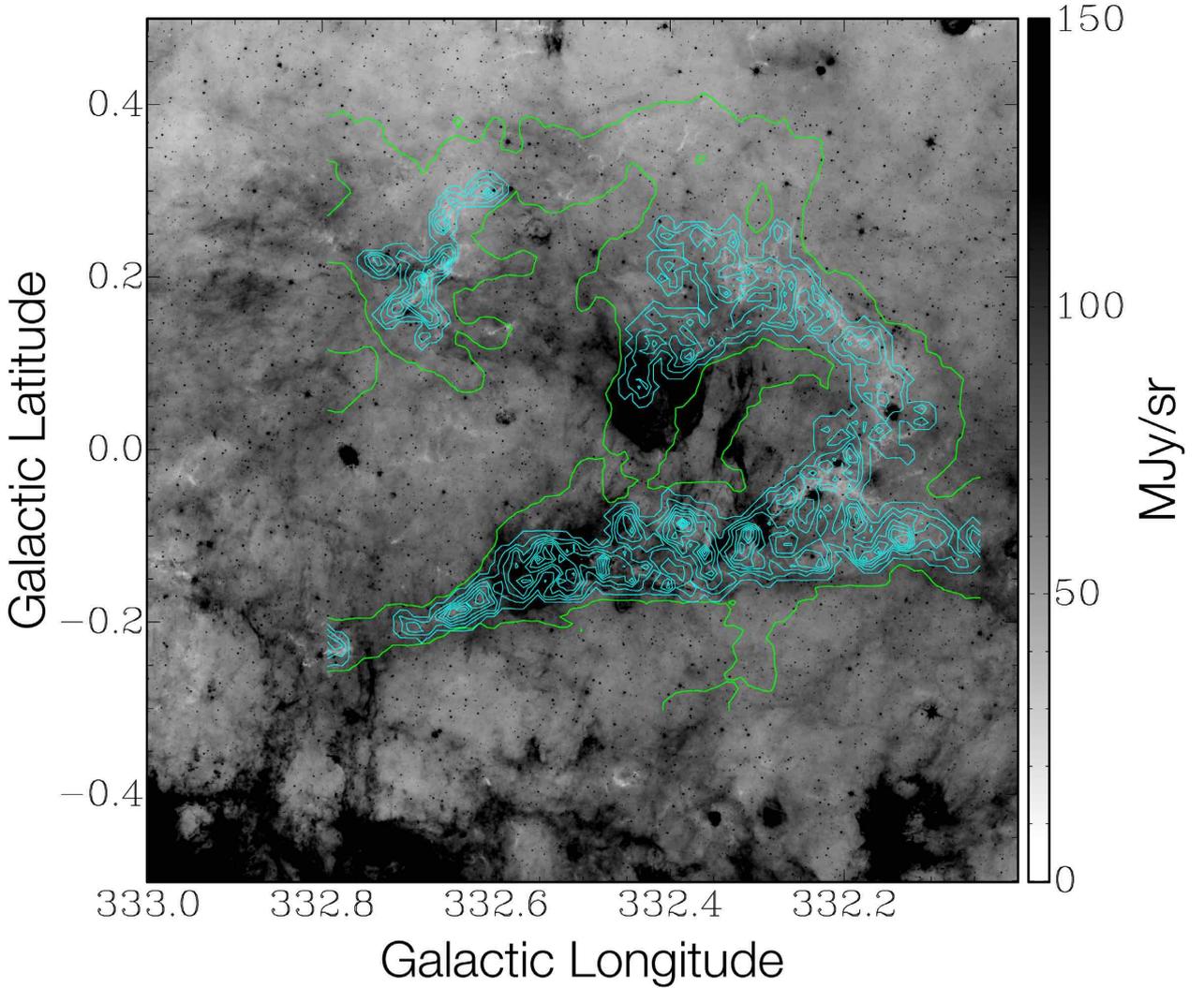

**Figure 32.** Spitzer 8 μm overlaid with the extent of the ring defined by $^{13}$CO (green contours) and in the cyan contours N(H$_2$) as traced by C$^{18}$O. Contour levels are from $5 \times 10^{20}$ to $6 \times 10^{22}$ in steps of $7 \times 10^{21}$ cm$^{-2}$.

We derived the X$_{co}$ variation along the ring for different T$_{ex}$ values, ranging from X$_{co}$ = 2.1 × 10$^{20}$ to 2.3 × 10$^{20}$ cm$^{-2}$ K$^{-1}$ km$^{-1}$ s, respectively, for a T$_{ex}$ = 10 K and for a T$_{ex}$ derived from peak temperature map of $^{12}$CO.

We defined an X$_{CI}^{809}$ factor, using the [C I] $^3P_2 \rightarrow$ $^3P_1$ transition at 809 GHz and N(H$_2$) traced by $^{13}$CO. The derived values are (1.77 ± 0.03) × 10$^{21}$, (1.83 ± 0.02) × 10$^{21}$, and (1.97 ± 0.02) × 10$^{21}$ cm$^{-2}$ K$^{-1}$ km$^{-1}$ s, respectively, for N(H$_2$) traced by the three assumptions of $^{12}$CO, $^{13}$CO without adding dark gas, and $^{13}$CO adding the dark gas fraction.

The median value of the $^{12}$CO/$^{13}$CO ratio is ∼3.3, with a minimum value close to 1, mostly in the [NW-R], region that we concluded to be the most quiescent part of the cloud. The median [C I]/$^{13}$CO ratio for the whole cloud is equal to 0.5, ranging in the interval [0.4–1], with the lowest values in [NW-R] and [W-R].

The [C I] emission is lowest in the [NW-R] and [W-R], while the maximum values are found to be in the Southern cloud regions.

The mean dark molecular gas mass fraction of N(H$_2$) not traced by CO results to be equal to ∼16 per cent, in average, considering a T$_{ex}$ = 20 K for $^{13}$CO, and a T$_{ex}$ = 30 K for [C I].

We compiled a catalogue of clumps decomposing the ring GMC C$^{18}$O emission and reporting the main physical characteristics derived from CO line, mostly all possess a virial parameter <1.

HiGAL maps were used to derive the dust SED and the N(H$_2$) traced by dust. The derived dust temperature is in the range between 16 and 27 K. The linear fit between N(H$_2$) traced by dust and the $^{13}$CO integrated intensity along all the line of sight gives a coefficient $m = 5.9 \times 10^{20}$ cm$^{-2}$ K$^{-1}$ km$^{-1}$ s and $b = 9.8 \times 10^{21}$ cm$^{-2}$.

We presented a method to disentangle dust emission across the spiral arms from our adopted Galactic model (McClure-Griffiths & Dickey 2007; Vallée 2014), through the use of the $^{13}$CO emission along the velocity range [−115, −30] km s$^{-1}$ (Table 6).

We found a decreasing gradient in Hi-Gal clumps temperatures and L$_{bol}$/M$_{clump}$ ratio going from [SE-R] towards [N-R], which corresponds to a T$_{ex}$ decrease and to an increase of X$_{co}$ median value and values dispersion.

The spatial distribution of cold dust and C$^{18}$O follows an arc-like 8 μm extinction feature in northern [SW-R], [W-R], [NW-R], and [E,SE-R]. We found correspondences between the lowest 8 μm emission, cold dust, and C$^{18}$O distribution also in the southern ring.

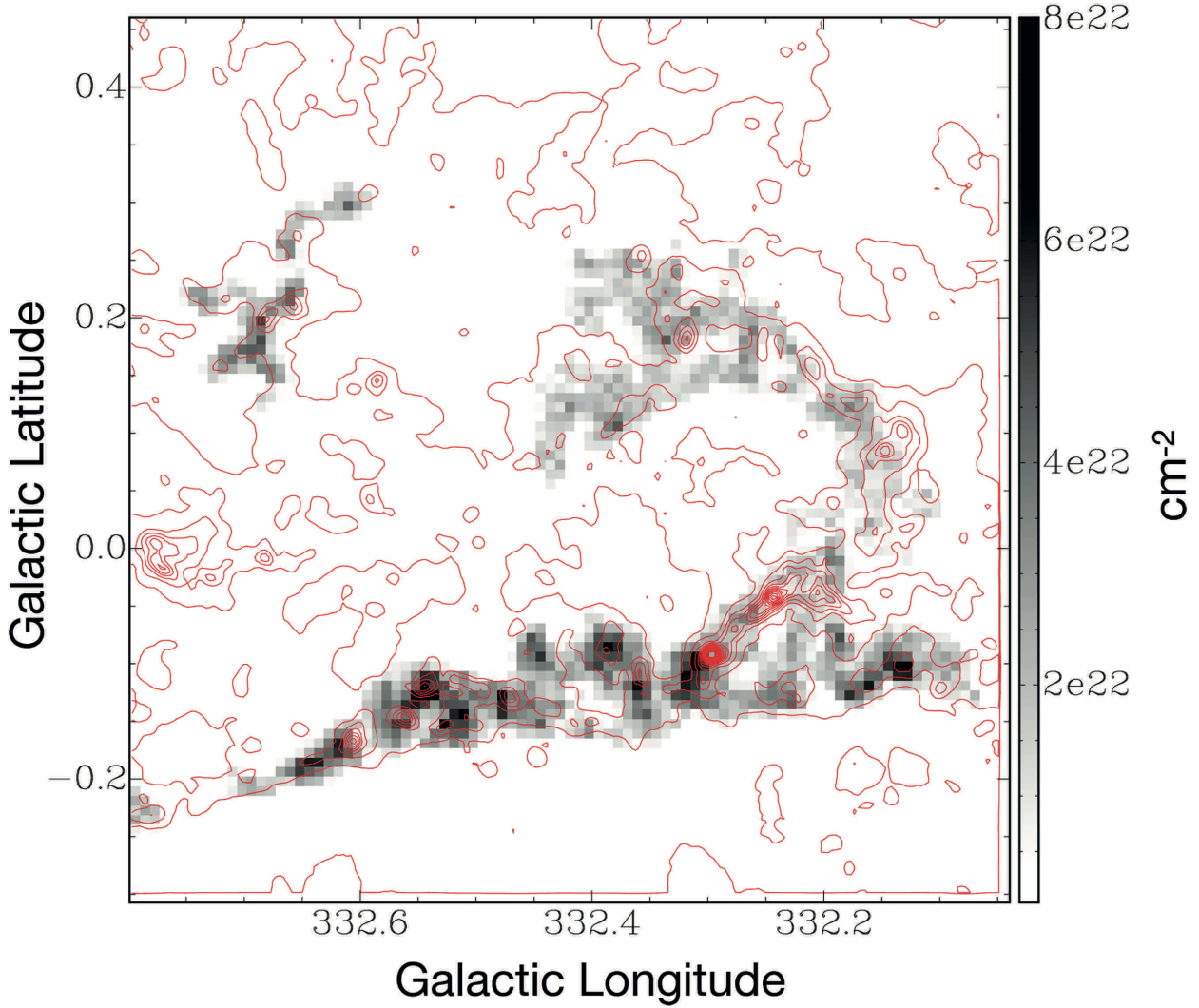

**Figure 33.** Molecular hydrogen column density from $C^{18}O$ and in red contours $N(H_2)$ traced by dust as per Fig. 13.

We consider this extinction feature as an IRDC that is inside the molecular ring. Along the IRDC we found that, in general, the peak position of $N(H_2)$ traced by $C^{18}O$ present an offset position with respect to $N(H_2)$ traced by dust peaks.

We concluded that the observed gas properties are influenced by the stellar formation activity occurring along the identified IRDC, which is an unique structure around which the molecular cloud is wrapped. The path from [E,SE-R] to Southern [N-R] is ∼65 pc. We found evidence of a different star formation activity between the southern and northern ring.

We presented a new visualization approach for the 3D radio astronomy datacubes by using VR and AR environment, based on UNITY, JAVASCRIPT, and online platforms in combination with non-specifically designed equipment.

## ACKNOWLEDGEMENTS

The Mopra radio telescope is part of the Australia Telescope National Facility. Operations support was provided by the University of New South Wales, the University of Adelaide, and Western Sydney University. Many staff of the ATNF have contributed to the success of the remote operations at Mopra. We particularly wish to acknowledge the contributions of Balthasar Indermuehle, David Brodrick, Philip Edwards, Brett Hiscock, and Peter Mirtschin.

Funding for the HEAT telescope is provided by the National Science Foundation under grant number PLR-0944335. PLATO-R was funded by Astronomy Australia Limited, as well as the University of New South Wales, as an initiative of the Australian Government being conducted as part of the Super Science Initiative and financed from the Education Investment Fund. Logistical support for HEAT and PLATO-R is provided by the United States Antarctic Program. We also acknowledge ARC support through Discovery Project DP120101585.

This research made use of ASTROPY, a community-developed core PYTHON package for Astronomy (Astropy Collaboration 2013) and the IDL ASTRONOMY LIBRARY (Landsman 1995), the SIMBAD data base (Wenger et al. 2000), and ALADIN SKY ATLAS desktop program (Bonnarel et al. 1999, 2000; Chilingarian & Zolotukhin 2008), maintained by the *CDS* in Strasbourg, France, and the GLUE[18] (Beaumont, Goodman & Greenfield 2015) visualization package. We thank the anonymous referee for the useful comments that helped us to enrich the work presented in this paper. We would also thank, for the stimulating discussions, Shaila Akhter, Nigel Maxted, Davide Elia, Matthew Freeman, Anant Tanna, and John Lopez.

[18]http://glue-viz.org

# APPENDIX A: HOW TO CREATE VR/AR CONTENT FROM FITS DATACUBES

An example of our VR approach is the $^{13}$CO ring emission at $3\sigma$ level, presented in a web page we created based on WEBVR,[19] using the WEBVR BOILERPLATE[20] framework. It requires an IOS or ANDROID device and a VR headset. We will now outline the three steps required to create the VR version of the $^{13}$CO ring emission.

First step: Convert the FITS format to another 3D format. Using a custom PYTHON routine based on the MAYAVI[21] package, we extracted a portion of the $^{13}$CO 3D FITS radio datacube and converted it to WRL[22] 3D format to make it compatible to other 3D software. A 3D mesh is then created applying a sigma threshold (in this case equal to $3\sigma$) on the extracted datacube section; this mesh is a connected surface (isosurface) that comprises all the voxels whose emission levels are above the threshold. It is important to note that the mesh grid is the same of the starting FITS datacube.

Second step: the mesh in WRL format is converted to JSON 3D format by using the 3D online editor CLARA.IO.[23] In this way it becomes readable by the JAVASCRIPT[24] library THREE.JS.[25] This step is necessary to allow the reading of the 3D model by the WEBVR BOILERPLATE, which uses THREE.JS library to create VR content.

Third step: web page creation, putting together HTML and JAVASCRIPT code to configure the WEBVR environment, linking and setting the 3D JSON model to THREE.JS JAVASCRIPT library. The final web page is at http://hologhost.altervista.org/VR/index.html. To experience the best view we recommend the use of a pair of VR headset and a smartphone or dedicated VR hardware. We tested it on both IOS and ANDROID devices.

We verified the coherency between the 3D CO emission distribution, as in the original FITS datacube, and the final mesh plotted by THREE.JS in WEBVR environment. Comparing the mesh grid spacing and the original emission distribution we found no discrepancy between the two representations.

The limit of this method is the impossibility to move inside the VR environment without using a dedicated controller. A solution could be to change the point of view coordinates, but this drastically decreases the interaction rapidity with the model. The advantages are its accessibility and the use of non-proprietary software.

An alternative we explored is based on the SKETCHFAB[26] online platform,[27] which gives the possibility to inspect 3D models both in VR or AR on mobile devices (both IOS or ANDROID with VR headset) or on desktop PC equipped with VR hardware. This solution is partially free, there is the possibility to choose between a basic or paid accounts, the major difference is on the size limits of the uploaded model. The advantage of this solution is the immediate creation of cross-platform VR and AR environments without any required coding. Here[28] there is an AR/VR example of the C$^{18}$O 3D clumps distribution as reported in Section 9 and in Table A1. The most interesting feature is the AR function accessible via mobile app. In other words it is possible to virtually place the object on a real detected surface and move around it as it is in the real world. Here[29] there is a fully navigable 3D representation (VR/AR mode) of the molecular ring $4\sigma$ emission coming from all the three CO isotopologues.

Another approach made use of the UNITY[30] engine to build AR IOS applications for both smartphones and tablets. Though giving details on the programming techniques used in this methodology is beyond the purpose of this paper, we report the principal steps through which we created an application usable on mobile devices alone or in combination with a mobile headset system.

---

[19] WEBVR is an open project to create VR content accessible via the web regardless of the used device; it requires a VR headset, an IOS, or ANDROID mobile device with accelerometers, and a compatible browser: https://webvr.info/.
[20] https://github.com/borismus/webvr-boilerplate, a framework to develop cross-platform (smartphone with VR headset or desktop VR systems) WEBVR applications.
[21] http://code.enthought.com/pages/mayavi-project.html.
[22] https://en.wikipedia.org/wiki/VRML
[23] https://clara.io/.
[24] https://en.wikipedia.org/wiki/Javascript.
[25] https://threejs.org/.
[26] https://sketchfab.com/feed
[27] At the moment of writing we note that it works on SAFARI and FIREFOX browsers but not on CHROME; more informations at https://security.googleblog.com/2017/09/chromes-plan-to-distrust-symantec.html.
[28] https://sketchfab.com/models/e07902733c9b44e7b3bc0b7197bcffe4
[29] https://sketchfab.com/models/eeb0f06659024f35baa18ba4cf11139b
[30] https://unity3d.com/

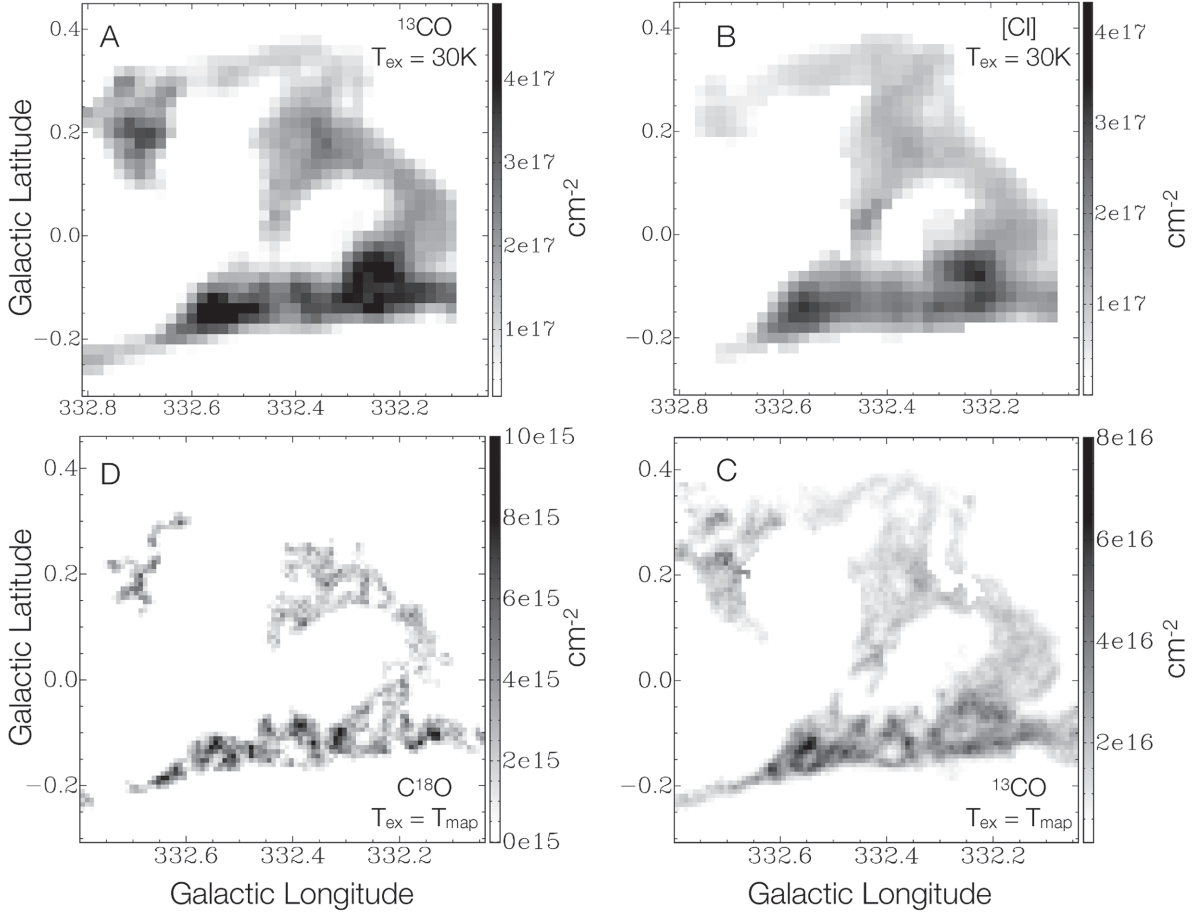

**Figure A1.** Column densities map for as indicated in the header. With A and B are indicated the column densities derived at a [C I] resolution for an assumed excitation temperature of 30 K. With D and C are indicated the column densities of $C^{18}O$ and $^{13}CO$, respectively, computed at the native CO resolution, assuming a $T_{ex}$ derived from the $^{12}CO$ peak temperature emission.

The first step is necessarily the conversion of the FITS datacube into a readable format for UNITY. We perform this step using the same custom PYTHON routine above mentioned then converting the mesh in OBJ[31] or DAE[32] 3D format. The datacube is thus transformed in a single (or multiple) mesh (depending on how the number of the visualized isosurfaces) ready to be imported in UNITY. The VR (and AR) capabilities we experimented with made use of the VUFORIA[33] framework, which offers the possibility to create virtual buttons, making possible to build virtual user interfaces through which we interact with the model. The porting on mobile device can be done in UNITY itself using the VUFORIA library, depending on the mobile operating system. As expected, we found that the increment in the complexity of the model dramatically slows the visualization experience on mobile devices, even though the meshes corresponding to the $3\sigma$ emission of $^{12}CO$, $^{13}CO$, and $C^{18}O$, coming from the G332 molecular ring, are easily displayed. Even though we tested this technology on mobile devices, a full immersion visualization experience can be achieved only with dedicated hardware, overcoming the limited computational constraints. On the other hand, the soft-ware tools complexity is continuously decreasing, making easier the creation of high-quality full immersion visualization solutions.

[31] http://paulbourke.net/dataformats/obj/
[32] https://www.khronos.org/collada/
[33] https://www.vuforia.com/

**Table A1.** Clumps found applying CLUMPFIND to the C$^{18}$O box smoothed datacube (see Section 5.1 for smooth parameters). In the first column the ID is followed by the flags *x*, *v* indicating that the clump partially extends, respectively, outside the Galactic Longitude interval of the ring or outside the ring velocity range, while *or* flags means that the clump is outside any of the ring regions (as defined in Fig. 1). The second, third, fourth, and fifth columns are for ℓ *b* clump centroid coordinates and the angular FWHMs of the corresponding clump ellipsoid, while $V_{LSR}$ is the velocity coordinate and ΔV is its FWHM. $R_c$ is the clump radius, used to derive $M_{vir}$ and $M_{LTE}^{20}$ under a constant spherical density profile assumption. $M_{LTE}^{20}$ is the clump mass computed using the method described in Section 6, for $T_{ex}$ = 20 K and considering C$^{18}$O emission. $α_{vir}$ is the virial parameter between $M_{vir}$ and $M_{LTE}^{20}$. $T_{AVG}^{18}$, $T_{AVG}^{13}$, $T_{12/13}^{AVG}$ and $T_{13/18}^{AVG}$ are, respectively, average $T_{mb}$ temperatures of the clump in the corresponding line and mean values of the relative lines ratio. $T_p^{18}$ is the clump peak brightness temperature in the C$^{18}$O line emission.

| | Centroid | | | | | C$^{18}$O clumps properties | | | | | | | | | |
|---|---|---|---|---|---|---|---|---|---|---|---|---|---|---|---|
| ID | ℓ | b | Δℓ (arcsec) | Δb (arcsec) | $V_{LSR}$ (km s$^{-1}$) | ΔV (km s$^{-1}$) | $R_c$ (pc) | $M_{vir}$ ($M_\odot$) | $M_{LTE}^{20}$ ($M_\odot$) | $α_{vir}$ | $T_{AVG}^{18}$ (K) | $T_{AVG}^{13}$ (K) | $T_{12/13}^{AVG}$ | $T_{13/18}^{AVG}$ | $T_p^{18}$ (K) |
| GMCC332.14−0.11$^x$ | 332.14° | −0.11° | 325.5 | 132.3 | −52.56 | 2.16 | 3.5 | 3440 | 11085 | 0.31 | 0.91 | 4.23 | 3.02 | 4.78 | 1.87 |
| GMCC332.19+0.11 | 332.19° | 0.11° | 217.8 | 279.3 | −49.98 | 1.63 | 3.3 | 1824 | 6526 | 0.28 | 0.84 | 4.00 | 2.48 | 4.85 | 1.51 |
| GMCC332.20−0.04 | 332.20° | −0.04° | 272.7 | 269.4 | −48.88 | 1.73 | 2.7 | 2420 | 8474 | 0.29 | 0.79 | 4.24 | 3.02 | 5.41 | 1.29 |
| GMCC332.24−0.13 | 332.24° | −0.13° | 259.8 | 195.6 | −50.72 | 2.05 | 3.2 | 3067 | 9816 | 0.31 | 0.82 | 4.00 | 2.99 | 4.90 | 1.27 |
| GMCC332.28+0.15 | 332.28° | 0.15° | 197.1 | 200.7 | −49.62 | 1.62 | 3.6 | 1652 | 5592 | 0.3 | 0.84 | 3.74 | 1.67 | 4.51 | 1.44 |
| GMCC332.29−0.01$^v$ | 332.29° | −0.01° | 42.3 | 71.7 | −44.83 | 0.63 | 0.9 | 74 | 236 | 0.32 | 0.73 | 2.42 | 2.79 | 3.33 | 0.87 |
| GMCC332.34+0.18 | 332.34° | 0.18° | 168.6 | 210.6 | −49.25 | 1.92 | 2.6 | 2045 | 5000 | 0.41 | 0.89 | 4.27 | 1.86 | 4.86 | 1.47 |
| GMCC332.36+0.20 | 332.36° | 0.20° | 229.8 | 200.7 | −49.25 | 1.91 | 3.9 | 2418 | 5734 | 0.42 | 0.83 | 3.92 | 2.58 | 4.73 | 1.27 |
| GMCC332.37−0.12 | 332.37° | −0.12° | 243.3 | 188.4 | −50.72 | 2.09 | 3.4 | 3159 | 13253 | 0.24 | 1.03 | 4.59 | 2.78 | 4.54 | 1.98 |
| GMCC332.39−0.09 | 332.39° | −0.09° | 135.6 | 119.4 | −49.62 | 1.37 | 2.3 | 917 | 5504 | 0.17 | 1.20 | 5.06 | 3.39 | 4.29 | 2.53 |
| GMCC332.39+0.11 | 332.39° | 0.11° | 247.5 | 199.5 | −48.14 | 1.39 | 3.4 | 1392 | 7179 | 0.19 | 0.93 | 4.20 | 2.92 | 4.61 | 1.73 |
| GMCC332.39+0.35 | 332.39° | 0.35° | 147.3 | 136.2 | −47.78 | 1.00 | 2.3 | 445 | 1580 | 0.28 | 0.77 | 3.42 | 3.10 | 4.47 | 1.07 |
| GMCC332.42+0.02 | 332.42° | 0.02° | 47.1 | 52.5 | −51.09 | 0.93 | 1.6 | 149 | 380 | 0.39 | 0.80 | 4.42 | 3.51 | 5.58 | 1.11 |
| GMCC332.45−0.10 | 332.45° | −0.10° | 221.4 | 240.6 | −50.35 | 2.50 | 3.6 | 4693 | 9867 | 0.48 | 0.90 | 4.41 | 3.55 | 4.98 | 1.47 |
| GMCC332.50+0.32 | 332.50° | 0.32° | 194.7 | 146.7 | −46.67 | 0.98 | 1.0 | 472 | 2112 | 0.22 | 0.77 | 3.03 | 2.90 | 3.99 | 1.07 |
| GMCC332.55−0.13 | 332.55° | −0.13° | 256.2 | 143.4 | −48.14 | 3.58 | 3.2 | 8571 | 11280 | 0.76 | 1.12 | 5.53 | 3.72 | 5.01 | 2.13 |
| GMCC332.59−0.13 | 332.59° | −0.13° | 186 | 167.4 | −48.88 | 2.92 | 3.0 | 4798 | 4024 | 1.19 | 0.85 | 4.74 | 4.07 | 5.57 | 1.33 |
| GMCC332.62+0.30$^o$ | 332.62° | 0.30° | 92.7 | 68.4 | −46.67 | 0.99 | 1.3 | 271 | 1227 | 0.22 | 0.96 | 4.10 | 2.89 | 4.39 | 1.55 |
| GMCC332.65−0.20 | 332.65° | −0.20° | 225.3 | 110.4 | −46.3 | 1.81 | 2.3 | 1599 | 5127 | 0.31 | 1.04 | 4.47 | 2.49 | 4.45 | 2.04 |
| GMCC332.67+0.20$^o$ | 332.67° | 0.20° | 64.5 | 35.4 | −47.41 | 0.89 | 0.8 | 138 | 355 | 0.39 | 0.78 | 3.96 | 1.35 | 5.15 | 0.93 |
| GMCC332.70+0.17$^o$ | 332.70° | 0.17° | 170.4 | 348.6 | −51.46 | 2.46 | 3.1 | 3981 | 7766 | 0.51 | 0.84 | 2.83 | 2.42 | 3.44 | 1.51 |
| GMCC332.74+0.21$^{v,o}$ | 332.74° | 0.21° | 132.6 | 88.8 | −44.83 | 1.49 | 3.5 | 755 | 1318 | 0.57 | 0.84 | 3.81 | 2.09 | 4.65 | 1.27 |
| GMCC332.74+0.31$^o$ | 332.74° | 0.31° | 93.3 | 91.8 | −45.2 | 0.89 | 1.4 | 237 | 716 | 0.33 | 0.74 | 3.79 | 2.83 | 5.09 | 0.96 |
| GMCC332.79+0.24$^o$ | 332.79° | 0.24° | 57.3 | 106.8 | −46.67 | 0.92 | 0.8 | 175 | 383 | 0.46 | 0.77 | 3.85 | 2.50 | 4.97 | 1.09 |
| GMCC332.80−0.23$^{x,v}$ | 332.80° | −0.23° | 102.9 | 82.2 | −44.46 | 1.40 | 1.4 | 574 | 1058 | 0.54 | 0.98 | 3.71 | 1.47 | 4.04 | 1.84 |

This paper has been typeset from a T$_E$X/L$^A$T$_E$X file prepared by the author.